\documentclass[11pt]{article}
\usepackage[utf8]{inputenc}
\usepackage{jheppub}
\usepackage{amsmath,amssymb,amsfonts,graphicx,slashed,color,amsthm,mathtools,upgreek, enumerate, tensor, scalerel, subfig}
\usepackage{arydshln}
\usepackage[dvipsnames]{xcolor}
\usepackage{bbm}
\allowdisplaybreaks

\title{Relational bulk reconstruction from modular flow
}

\author[a]{\!Onkar Parrikar,} 
\author[a]{\! Harshit Rajgadia,}
\author[a]{\! Vivek Singh,}
\author[b]{\! Jonathan Sorce}
\affiliation[\,a]{Department of Theoretical Physics, Tata Institute for Fundamental Research, 1 Homi Bhabha Road, Mumbai, Maharashtra 400005, India.}
\affiliation[\,b]{Center for Theoretical Physics, Massachusetts Institute of Technology, 182 Memorial Drive, Cambridge, MA 02142, USA.}
\newcommand{\ch}{\chi}
\newcommand{\tch}{\bar{\chi}}
\newcommand{\st}{\star}
\newcommand{\beq}{\begin{equation}}
\newcommand{\eeq}{\end{equation}}
\newcommand{\beqn}{\begin{eqnarray}}
\newcommand{\eeqn}{\end{eqnarray}}
\newcommand{\pa}{\partial}

\newcommand{\cH}{\mathcal{H}}

\newcommand{\cU}{\mathcal{R}}
\newcommand{\cA}{\mathcal{A}}

\newcommand{\cl}{\lambda}

\newcommand{\bra}[1]{\langle #1 \vert}
\newcommand{\ket}[1]{\vert #1 \rangle}
\newcommand{\inner}[2]{\langle #1 \vert #2 \rangle}

\DeclareMathOperator{\Tr}{Tr}

\newcommand{\cR}{\mathcal{R}}

\newcommand{\Abar}{{\overline{A}}}

\newcommand{\alg}{\mathcal{M}}

\renewcommand{\bar}{\overline}

\begin{document}

\abstract{
The entanglement wedge reconstruction paradigm in AdS/CFT states that for a bulk qudit within the entanglement wedge of a boundary subregion $\bar{A}$, operators acting on the bulk qudit can be reconstructed as CFT operators on $\bar{A}$. This naturally fits within the framework of quantum error correction, with the CFT states containing the bulk qudit forming a code protected against the erasure of the boundary subregion $A$. In this paper, we set up and study a framework for relational bulk reconstruction in holography: given two code subspaces both protected against erasure of the boundary region $A$, the goal is to relate the operator reconstructions between the two spaces. To accomplish this, we assume that the two code subspaces are smoothly connected by a one-parameter family of codes all protected against the erasure of $A$, and that the maximally-entangled states on these codes are all full-rank. We argue that such code subspaces can naturally be constructed in holography in a ``measurement-based'' setting. In this setting, we derive a flow equation for the operator reconstruction of a fixed code subspace operator using modular theory which can, in principle, be integrated to relate the reconstructed operators all along the flow. We observe a striking resemblance between our formulas for relational bulk reconstruction and the infinite-time limit of Connes cocycle flow, and take some steps towards making this connection more rigorous. We also provide alternative derivations of our reconstruction formulas in terms of a canonical reconstruction map we call the modular reflection operator.
}

\maketitle

\parskip=5pt

\section{Introduction}

The Ryu-Takayanagi formula \cite{Ryu:2006bv, Hubeny:2007xt, faulkner2013quantum, Engelhardt:2014gca} implies that the bulk-to-boundary map in AdS/CFT exhibits subregion-subregion duality \cite{Almheiri:2014lwa}.
Given a boundary subregion $\bar{A}$, there exists a corresponding bulk subregion $\bar{a}$ such that bulk degrees of freedom in $\bar{a}$ are encoded in, and can be reconstructed from, the boundary subregion $\bar{A}$.
The bulk region $\bar{a}$ is defined via an extremization procedure over codimension-2 bulk surfaces which are anchored on $\pa \bar{A}$, and is called the entanglement wedge \cite{Headrick:2014cta}.\footnote{In some cases, when there exist large discrepancies between so-called one-shot and asymptotic entanglement measures, a more refined classification of subregion-subregion duality is given in terms of ``max'' and ``min'' wedges \cite{Akers:2020pmf, Akers:2021fut, Akers:2023fqr}.}

The general arguments for subregion-subregion duality combine ideas from quantum error correction \cite{Papadodimas:2012aq, Dong:2016eik, Harlow:2016vwg, Cotler:2017erl}, semiclassical general relativity \cite{Headrick:2007km, Wall:maximin, Engelhardt:2014gca, Jafferis:JLMS}, and the gravitational path integral \cite{lewkowycz2013generalized, faulkner2013quantum, Dong:2016hjy, penington2022replica, almheiri2020replica}.
In the special case of a bulk operator lying in the causal wedge of the boundary region, there exists a simple formula for its boundary dual.
This is called the HKLL formula, and is expressed in terms of the bulk-to-boundary Green's functions for the bulk equations of motion \cite{Hamilton:2005ju, Hamilton:2006az}.
However, there exist many situations where the entanglement wedge is strictly larger than the causal wedge, and in this setting there is no general, manifestly Lorentzian formula for expressing bulk operators in terms of boundary operators.
There is a universal formula in terms of the Petz map \cite{Cotler:2017erl} that can be used to prove theorems about bulk reconstruction, but applications of this formula in AdS/CFT often involve the Euclidean path integral, are thus not manifestly Lorentzian, and are often impractical.\footnote{The Petz map is a quantum channel, and so must admit a Lorentzian description, potentially involving ancillary degrees of freedom. However, this Lorentzian description is not very manifest. See however \cite{penington2022replica, Vardhan:petz} for cases where the Petz map can be computed explicitly.}

One practical, Lorentzian method for reconstructing an operator that lies outside the causal wedge is to act on the boundary with sources that modify the spacetime so that the causal wedge expands to contain the operator in question.
In the modified spacetime, a bulk reconstruction formula is provided by HKLL, and conjugating the HKLL reconstruction map by the ``causal wedge expansion map'' produces a boundary formula for operators outside the causal wedge.
This was explored at the level of perturbation theory in \cite{Almheiri:2017fbd, Levine:2020upy}, and it was argued generally in \cite{Engelhardt:2021mue} that by iterating backward and forward boundary time evolution and inserting simple sources, one can always expand the causal wedge to the ``simple wedge'', i.e., the region of the entanglement wedge lying between the outermost extremal surface and the asymptotic boundary. It was also argued in \cite{Engelhardt:2021mue} that simple sources \emph{cannot} be used to expand the causal wedge any further than the simple wedge, so a stronger algorithm is needed to reconstruct operators in the region behind the outermost extremal surface.
The region behind the outermost extremal surface is called the \textit{python's lunch} \cite{Engelhardt:2017aux, Engelhardt:2018kcs, Brown:2019rox, Engelhardt:2021mue, Akers:2022qdl}.
Note that this method for bulk reconstruction explicitly relies on backreaction of the matter sources on spacetime.

A different approach to Lorentzian bulk reconstruction was proposed in \cite{Faulkner:zero-modes}, where the authors suggested that modular flow can be used as a tool to reconstruct operators beyond the causal wedge.
Unlike in the previous paragraph, reconstruction by modular flow is performed in a regime where there is no backreaction, and does not involve any modification of the spacetime.
Instead, it was conjectured in \cite{Faulkner:zero-modes} that by acting with modular flow on operators in the causal wedge, one can produce an arbitrarily good approximation to any operator in the entanglement wedge.
Since causal wedge operators can be reconstructed using the HKLL formula, the combination of HKLL and modular flow could then produce formulas for bulk reconstruction in the full entanglement wedge. 
In \cite{Faulkner:zero-modes}, this general conjecture was supported by an explicit formula for operators localized at the extremal surface.\footnote{See also \cite{Chen:island, Gao:local-modflow-reconstruction} for related work.}

A unifying feature of the above approaches to bulk reconstruction is that they are \emph{relational} in nature.
Given an operator for which there is no known reconstruction formula, one tries to express it in terms of operators for which reconstruction formulas are known.
The purpose of this paper is to develop a general framework for \emph{relational bulk reconstruction} in the absence of backreaction, by expressing one reconstruction map in terms of another, and to partially bridge the gap between the Petz map and the modular flow approaches to bulk reconstruction.
Our analysis rests on fundamental features of finite-dimensional quantum error-correcting codes, and invokes a ``measurement-based'' notion of what it means to reconstruct a bulk operator.

We now give a summary of our main results. In the most general terms, we consider finite-dimensional code spaces $\cH_{\text{code}},$ which are subspaces of a larger Hilbert space $\cH_{\text{phys}}$ which decomposes into systems $A$ and $\bar{A}.$
We assume that the code subspace $\cH_{\text{code}}$ describes excitations localized within $\bar{A},$ so that $\cH_{\text{code}}$ is protected against erasures of the region $A$.
A general phenomenon known as the decoupling principle --- see e.g. \cite{Schumacher_1996, Nielsen_1998, NielsenPoulin, Preskill:2016htv, Harlow:2016vwg, Balasubramanian:2022fiy}  --- tells us that protection against the erasure of $A$ is equivalent to the statement that if $\cH_{R}$ is a reference system of the same dimension as $\cH_{\text{code}},$ and $\ket{\Psi}$ is any maximally entangled state between $\cH_{\text{code}}$ and $\cH_{R},$ then $\ket{\Psi}$ decouples between the systems $A$ and $R$:
\begin{equation}
	\bra{\Psi} O_{A} \otimes O_{R} \ket{\Psi}
		= \bra{\Psi} O_{A} \ket{\Psi} \bra{\Psi} O_R \ket{\Psi}.
\end{equation}
Furthermore, if $\ket{\psi_1}, \dots, \ket{\psi_d}$ is an orthonormal basis for $\cH_{\text{code}},$ and $\phi_{jk}$ is a $d \times d$ matrix, then the error correction task of finding an operator $O_{\bar{A}}$ in $\bar{A}$ which acts on $\cH_{\text{code}}$ in the same way as the global operator $\sum_{jk} \phi_{jk} \ket{\psi_j} \bra{\psi_k}$ is the same as solving the equation
\begin{equation} \label{eq:transpose-trick-equation}
	O_{\bar{A}} \ket{\Psi}= \sum_{j,k} \phi_{kj} \ket{k}  \inner{j}{\Psi},
\end{equation}
where $\ket{j}$ denotes the orthonormal basis of the reference system that is paired with the basis $\ket{\psi_j}$ in the maximally entangled state $\ket{\Psi}.$ 
These observations motivate the study of maximally entangled states between a code and a reference --- sometimes called Choi-Jamiolkowski states --- as fundamental objects in quantum error correction.

In section \ref{sec:main-section}, we introduce our framework for relational bulk reconstruction. We consider two code subspaces $\cH_{\text{code}}(0)$ and $\cH_{\text{code}}(1)$, both exactly protected against erasure of $A$, with the goal of relating the reconstruction map of the latter code to the former. In other words, given a fixed set of matrix elements on the code subspace, and given operators $O_{\bar{A}}(0)$ and $O_{\bar{A}}(1)$ that reproduce those matrix elements on $\cH_{\text{code}}(0)$ and $\cH_{\text{code}}(1)$ respectively, we would ideally like to find a formula of the form:
\beq 
O_{\bar{A}}(1) = U_{1|0}^{\dagger} \,O_{\bar{A}}(0)\, U_{1|0},
\eeq 
for some unitary operator $U_{1|0}$ localized on the region $\bar{A}$. To this end, we assume that we can find a one-parameter family of code subspaces $\cH_{\text{code}}(\lambda)$ interpolating between $\cH_{\text{code}}(0)$ and $\cH_{\text{code}}(1),$ and an associated one-parameter family of maximally entangled states $\ket{\Psi_{\lambda}}.$ 
We assume that this one-parameter family of codes is exactly protected against erasure of $A$, and further, that the maximally entangled states $\ket{\Psi_{\lambda}}$ are cyclic and separating for the subsystem $\bar{A}$, which is the continuum version of the state having full Schmidt rank for the bipartition $\bar{A}:AR$.  With these assumptions, we derive a flow equation for $\ket{\Psi_{\lambda}},$ given by
\begin{equation} \label{eq:state-flow-intro}
	\ket{\dot{\Psi}}
		=  \ket{\delta \Psi} + \lim_{\epsilon \to 0^+} i \int_0^{\infty} dt\, e^{- \epsilon t} \Delta^{-it} \dot{h} \ket{\Psi} ,
\end{equation}
where $\Delta$ is the modular operator of the state $\ket{\Psi}$ with respect to the subsystem $\bar{A},$ $h = - \log \Delta$ is the full modular Hamiltonian, and $\ket{\delta \Psi}$ is the projection of $|\dot{\Psi}\rangle$ onto the space of ``modular zero modes,'' i.e., onto the zero-eigenspace of $h$.
Equation \eqref{eq:transpose-trick-equation} together with decoupling then implies that the reconstructed operators on $\bar{A}$ satisfy the flow equation
\begin{equation} \label{eq:operator-flow-intro}
	\dot{O}_{\bar{A}}
		= [\widehat{a}, O_{\bar{A}}] + i \int_0^{\infty} dt\, e^{- \epsilon t} \Delta^{-it} [\dot{h}, O_{\bar{A}}] \Delta^{it},
\end{equation}
where we have left the $\epsilon$-limit implicit, and where $\widehat{a}$ is an operator in $\bar{A}$ that is left invariant by modular flow and which is related to the term $|\delta \Psi\rangle$ in equation \eqref{eq:state-flow-intro}.
We also argue that in many settings of interest, the first term of equation \eqref{eq:operator-flow-intro} can be dropped, resulting in an equation that depends only on modular flow.
If the $\epsilon$ regulator in equation \eqref{eq:operator-flow-intro} is treated naively, then the integral in equation \eqref{eq:operator-flow-intro} can be expressed in terms of an infinite-time limit of Connes cocycle flow between the states $\ket{\Psi_0}$ and $\ket{\Psi_1}$, resembling previous work on the Connes cocycle flow in \cite{Levine:2020upy, Lashkari:2019ixo}.
In section \ref{sec:connes-cocycle-connection}, we outline a more rigorous argument for connecting equation \eqref{eq:operator-flow-intro} to the Connes cocycle, although we leave a detailed analysis for future work. 
We emphasize that equation \eqref{eq:operator-flow-intro} is a fully Lorentzian equation that can be integrated to perform relational reconstruction using modular flow for codes satisfying the cyclic-separating property for maximally entangled states.

In section \ref{sec:full-rank-codes-holography}, we investigate the assumption that the maximally entangled states $\ket{\Psi_{\lambda}}$ are cyclic and separating for the reconstructible region $\bar{A}$.
The validity of this assumption is one of the key issues governing whether equation \eqref{eq:operator-flow-intro} can be used for bulk reconstruction in holography.
Because we are using tools from exact quantum error correction, we are implicitly working in the infinite-$N$ regime of holography, as bulk reconstruction is known to be approximate at finite $N$ \cite{Cotler:2017erl, Hayden:BH-alpha-bits}.
It turns out that it is not natural for a holographic code to have a cyclic-separating maximally entangled state.
However, we argue there is a natural class of codes at infinite $N$ which do possess the cyclic-separating property, and which are relevant for studying AdS/CFT.
These are subspaces of an enlarged Hilbert space $\cH_{\text{CFT}} \otimes \cH_{Q},$ where $\cH_{Q}$ is a qudit used to probe excitations around a fixed holographic state $\ket{\psi} \in \cH_{\text{CFT}}$.
The code subspaces we consider consist of the states
\begin{equation} \label{eq:measurement-codes-intro}
	U_{\bar{a}, Q} \left(\ket{\psi}_{\text{CFT}} \otimes \ket{\chi}_{Q} \right),
\end{equation}
where $\ket{\chi}_{Q}$ is an arbitrary state of $\cH_{Q},$ and $U_{\bar{a}, Q}$ is a fixed unitary that couples the auxiliary system $Q$ to a local operator in the entanglement wedge $\bar{a}$ of $\bar{A}$ that one wishes to probe.
One can heuristically think of these states as being prepared by bringing in an external, auxiliary qudit and ``placing it'' in the holographic spacetime via the coupling $U_{\bar{a}, Q}$.
As we explain in section \ref{sec:full-rank-codes-holography}, while operators acting on the code subspace \eqref{eq:measurement-codes-intro} are not themselves operators in $\cH_{\text{CFT}},$ they can be used to perform arbitrary measurements on bulk operators in the entanglement wedge of $\bar{A},$ which suffices for all practical purposes.
This resembles the work of \cite{Lamprou:probes1, Lamprou:probes2, Lamprou:probes3}, in which it was shown that certain features of a holographic black hole can be measured by coupling it to an auxiliary black hole which acts as a probe.

In section \ref{sec:modular-reflections}, we give an alternative derivation of the fundamental equation \eqref{eq:operator-flow-intro} using a reconstruction map associated with the \emph{modular reflection operator}.
The modular reflection operator is associated with a canonical purification of a maximally entangled state on the code subspace; it was introduced in \cite{Parrikar:quantum-extremal-shock}, and was inspired by the application of canonical purifications to holography in \cite{Engelhardt:2017aux, Engelhardt:2018kcs, Dutta:canonical-purification}, together with the ideas in \cite{Harlow:2016vwg}.
We explain a natural connection between the modular reflection operator and the decoupling principle, and comment on the potential utility of the modular reflection operator for extending equation \eqref{eq:operator-flow-intro} beyond the regime of measurement operators at infinite $N$.

Before proceeding to the main body of the paper, we wish to emphasize three points.
The first is that our relational formula \eqref{eq:operator-flow-intro} can be used in principle to probe a bulk qudit arbitrarily deep in a non-island region of the entanglement wedge, by deforming the qudit through a one-parameter family of codes corresponding to bulk qudits placed at different locations, and eventually to a location inside the causal wedge where a reconstruction map is furnished by HKLL.
The second point is that equation \eqref{eq:operator-flow-intro} provides a realization of the proposal of \cite{Faulkner:zero-modes} that modular flow can be used to reconstruct operators behind the causal wedge, with the caveat that we are reconstructing measurement operators that probe a finite-dimensional code subspace of the bulk, instead of reconstructing operators in the full, infinite-dimensional Hilbert space of bulk quantum field theory.
The final point we wish to emphasize is that since our investigations take place in the infinite-$N$ limit of a holographic field theory, there is no backreaction, meaning that our work is not a direct generalization of \cite{Almheiri:2017fbd, Levine:2020upy, Engelhardt:2021mue}.
We think it would be very interesting to see if our approach can be applied to the scenarios considered in \cite{Almheiri:2017fbd, Levine:2020upy, Engelhardt:2021mue}.

\section{Relational reconstruction for full-rank codes}
\label{sec:main-section}

In subsection \ref{sec:full-rank-state-flow}, we derive the fundamental flow equation \eqref{eq:state-flow-intro} for a one-parameter family of cyclic and separating states.
In subsection \ref{sec:full-rank-code-flow}, we apply this equation to quantum error correcting codes with a cyclic-separating maximally entangled state, and derive equation \eqref{eq:operator-flow-intro}; we also argue that the first term in that equation will vanish in many holographic applications. 
Finally, in subsection \ref{sec:connes-cocycle-connection}, we discuss an apparent connection between our flow equations and the infinite-time limit of the Connes cocycle.

All of the necessary statements from modular theory are stated explicitly, and reading this section should not require any foreknowledge of modular flow.
However, see \cite{Sorce:2023fdx} for a review of von Neumann algebras, and \cite{Sorce:2023gio} for a physical explanation of modular flow.

\subsection{Flow equations for full-rank states}
\label{sec:full-rank-state-flow}

Consider a tensor product Hilbert space, $\cH = \cH_A \otimes \cH_{\bar{A}}.$
Every state $\ket{\Psi} \in \cH$ admits a Schmidt decomposition
\begin{equation}
	\ket{\Psi} = \sum_{j} \sqrt{p_j} | \chi_{j} \rangle\otimes  |\bar{\chi}_j \rangle,
\end{equation}
where the numbers $p_j$ are nonnegative, $| \chi_{j} \rangle$ is an orthonormal basis for $\cH_A$, and $| \bar{\chi}_j \rangle$ is an orthonormal basis for $\bar{A}.$
The state is said to have \textit{full Schmidt rank} if none of the numbers $p_j$ are zero.
This is easily seen to be equivalent to the statement that if an operator $O_A$ acting on $\cH_A$ annihilates $\ket{\Psi},$ then we have $O_A = 0,$ and similarly for an operator $O_{\bar{A}}$ acting on $\cH_{\bar{A}}.$
Any state $\ket{\Psi}$ with the property
\begin{equation}
	O_A |\Psi\rangle = 0,\;\;\; \Rightarrow\;\;\; O_A = 0,
\end{equation}
is called \textit{separating for $A$}.
Equivalently, it is called \textit{cyclic for $\bar{A}$}.
It is a straightforward exercise to show that a state is cyclic for $\bar{A}$ if and only if the set $\{O_{\bar{A}} \ket{\Psi}\}$ is dense in $\cH$.

In holography, one often considers the division of a conformal field theory into complementary spacetime regions $A$ and $\bar{A}$.
In this setting, the CFT Hilbert space $\cH_{\text{CFT}}$ does not actually factorize into a Hilbert space corresponding to $A$ and a Hilbert space corresponding to $\bar{A}$; instead, there is a von Neumann algebra $\alg$ of field operators acting on $A$, and a von Neumann algebra $\bar{\alg}$ of field operators acting on $\bar{A}$.
Typically, it is assumed that $\alg$ and $\bar{\alg}$ are each other's commutants, so that an operator is in $\alg$ if and only if it commutes with every operator in $\bar{\alg},$ and vice versa.
While Schmidt decompositions no longer make sense, we can still say that a state $\ket{\Psi} \in \cH_{\text{CFT}}$ is separating for $A$ if the only operator in $\alg$ annihilating $\ket{\Psi}$ is the zero operator; we can say it is cyclic for $A$ if the space $\alg \ket{\Psi}$ is dense in $\cH_{\text{CFT}}$.
In this continuum setting, cyclic and separating states play the same role as states that are full-rank across a chosen bipartition.
It is known (see e.g. \cite{Witten:2018zxz}) that states of bounded energy-momentum in quantum field theory are cyclic and separating for any division of Minkowski spacetime into complementary regions, and in more general spacetimes there is expected to be an abundance of cyclic and separating states \cite{Strohmaier:reeh, Sanders:reeh, Strohmaier:timelike-tube}.

Generally speaking, it is not necessary to work in the algebraic setting to study quantum field theories.
One can simply assume that the Hilbert space factorizes --- since this is the case when the field theory is regulated on a lattice, anyway --- and obtain all the same physical answers as would be obtained by working directly in the continuum.
Indeed, the flow equations we derive in this subsection are rederived in section \ref{sec:modular-reflections} using the assumption that the Hilbert space factorizes.
However, as we will see, it is actually more straightforward to derive a flow equation by working in the algebraic setting, since this is the setting in which the modular operator is typically studied.

Given a Hilbert space $\cH$ with a von Neumann algebra $\bar{\alg},$ and given a state $\ket{\Psi}$ that is cyclic and separating for $\bar{\alg},$ one can always define an associated modular operator $\Delta$ and a modular Hamiltonian $h = - \log\Delta.$
The modular operator $\Delta$ is a positive, self-adjoint, generally unbounded operator, and the associated unitary group $\Delta^{-it}$ is called the modular flow.
It is the continuum generalization of the operator
\begin{equation} \label{eq:density-matrix-modular}
	\rho_{\alg}^{-1} \otimes \rho_{\bar{\alg}},
\end{equation}
which can be defined in the factorized setting.
The modular flow satisfies three nice properties: (i) it fixes the state ($\Delta^{-it} \ket{\Psi} = \ket{\Psi}$); (ii) it fixes the algebras ($\Delta^{-it} \alg \Delta^{it} = \alg,\; \Delta^{-it} \bar{\alg} \Delta^{it} = \bar{\alg}$), and (iii) it admits a well behaved analytic continuation in $t$ governed by a technical property known as the KMS condition.

We will now derive equation \eqref{eq:state-flow-intro} using the basic tools of modular theory.
Note that because modular operators are unbounded, there are some subtleties that must be dealt with to manipulate them rigorously.
For example, if we have a one-parameter family of modular Hamiltonians $h(\lambda),$ and want to define the derivative
\begin{equation}
	\dot{h}(\lambda)
	= \lim_{\epsilon \to 0} \frac{h(\lambda + \epsilon) - h(\lambda)}{\epsilon},
\end{equation}
then we must be careful about the sense in which we expect the limit to exist, since $h(\lambda+\epsilon)$ and $h(\lambda)$ might not even act on the same subspace of Hilbert space.
In the following, we will ignore such subtleties, and derive equation \eqref{eq:state-flow-intro} using formal manipulations where all limits are assumed to be well behaved.\footnote{It is often the case that unjustified manipulations of an unbounded operator $h$ can be made mathematically rigorous by rewriting them as manipulations of the bounded operator $e^{ih}$; see for example section \ref{sec:connes-cocycle-connection}, where manipulations of this kind are used to derive a version of equation \eqref{eq:state-flow-intro} in which the Connes cocycle appears explicitly.}

A state $\ket{\Psi}$ with modular operator $\Delta$ satisfies the equation
\begin{equation}
(\Delta^{1/2} - 1) \ket{\Psi} = 0.
\end{equation}
Now let $|\Psi_{\lambda}\rangle$ be a one-parameter family of cyclic-separating states.
At a given value of $\lambda,$ let $|\chi\rangle$ be a fixed vector in the domain of $\Delta^{1/2}.$
Taking an overlap of the above equation with $|\chi\rangle,$ we obtain
\begin{equation}
	0 = \langle \chi|(\Delta^{1/2}-1)|\Psi\rangle= \int_0^{1/2}d\tau\,\langle \chi|\pa_{\tau}\Delta^{\tau}|\Psi\rangle= - \int_0^{1/2} d\tau\, \langle \Delta^{\tau} \chi | h \Psi \rangle.
\end{equation}
Taking a derivative with respect to $\lambda,$ using the identity $h \ket{\Psi} = 0,$ and denoting $\lambda$ derivatives with an overdot, we obtain
\begin{equation}
	\langle (\Delta^{1/2} - 1) \chi | \dot{\Psi} \rangle = \int_0^{1/2} d\tau\, \langle \Delta^{\tau} \chi | \dot{h} \Psi \rangle.
\end{equation}
Since $|\chi\rangle$ is in the domain of $\Delta^{1/2}$, the bra-valued function $z \mapsto \langle \Delta^{\bar{z}} \chi |$ is holomorphic in the strip $0 \leq \text{Re}(z) \leq 1/2$ --- see for example \cite[section 2.4]{Sorce:2023gio}.
Consequently, the contour in the above equation can be deformed to the vertical rays $0 - i t$ and $1/2 - i t$, together with a contribution along the segment $[0, 1/2] - i \infty$.
To ensure that the contribution at infinity can be dropped, we introduce a regulator before deforming the contour, writing
\begin{equation}
	\langle (\Delta^{1/2} - 1) \chi | \dot{\Psi} \rangle
	= \int_0^{1/2} d\tau\, e^{-i \epsilon \tau} \langle \Delta^{\tau} \chi | \dot{h} \Psi \rangle,
\end{equation}
and leaving the limit $\epsilon \to 0^+$ implicit.
After deforming the contour, we may now drop the term at infinity to obtain
\begin{equation}
	\langle (\Delta^{1/2} - 1) \chi | \dot{\Psi} \rangle
	= (-i) \int_0^{\infty} d t\, e^{- \epsilon t} \langle \chi | \Delta^{- i t}\dot{h} |\Psi\rangle
	+ i \int_0^{\infty} d t\, e^{- \epsilon t} e^{- i \epsilon/2} \langle \Delta^{1/2}  \chi | \Delta^{- i t}\dot{h} \ket{\Psi}.
\end{equation}
The factor $e^{- i \epsilon/2}$ does not matter in the $\epsilon \to 0$ limit, so we may neglect it to obtain
\begin{equation}
	\langle (\Delta^{1/2} - 1) \chi |\dot \Psi\rangle = \langle (\Delta^{1/2} - 1) \chi | \int_0^{\infty} d t\, i e^{- \epsilon t} \Delta^{- i t}\dot{h} |\Psi\rangle.
\end{equation}
From this equation, we conclude that $\ket{\dot{\Psi}}$ may be written as
\begin{equation}
	|\dot \Psi\rangle = \ket{\delta \Psi} + i \int_0^{\infty} d t\, e^{- \epsilon t} \Delta^{-i t}\dot{h} |\Psi\rangle,
\end{equation}
where $|\delta \Psi\rangle$ is orthogonal to the image of $(\Delta^{1/2} - 1).$
This implies $\Delta^{1/2} \ket{\delta \Psi} = \ket{\delta \Psi}.$
In fact, the second term in this equation is orthogonal to the trivial eigenspace of $\Delta^{1/2}$, since for any $\ket{\chi}$ satisfying $\Delta^{1/2} \ket{\chi} = \ket{\chi},$ we have
\begin{equation}
	\int_0^{\infty} d t\, e^{- \epsilon t} \bra{\chi} \Delta^{-i t}\dot{h} |\Psi\rangle
		= \int_0^{\infty} d t\, e^{- \epsilon t} \bra{\chi} \dot{h} |\Psi\rangle
		= \frac{1}{\epsilon} \bra{\chi} \dot{h} \ket{\Psi},
\end{equation}
and this vanishes by differentiating the equation $\bra{\chi} h \ket{\Psi} = 0$ and using $h \ket{\chi} = h \ket{\Psi} = 0.$
We have therefore reproduced equation \eqref{eq:state-flow-intro}, together with the observation that the first term is not just in the zero-eigenspace of $h,$ but is in fact the projection of $\ket{\dot{\Psi}}$ onto the zero-eigenspace of $h$.

It is a general fact about modular operators --- see for example \cite[appendix B]{Sorce:2023gio} --- that any state fixed by $\Delta^{1/2}$ can be written as $\ket{\delta \Psi} = \widehat{a} \ket{\Psi},$ where $\widehat{a}$ is an operator affiliated with $\bar{\alg}$ and satisfying $\Delta^{it} \widehat{a} \Delta^{-it} = \widehat{a}.$
The term ``affiliated'' means that $\widehat{a}$ commutes with every operator in $\alg$; strictly speaking, we cannot say that $\widehat{a}$ is ``in'' $\bar{\alg}$ because it may be unbounded, and $\bar{\alg}$ contains only bounded operators.
This allows us to write the flow equation as 
\begin{equation} \label{eq:state-flow-with-zero-mode}
	|\dot \Psi\rangle = \widehat{a} \ket{\Psi} + i \int_0^{\infty} d t\, e^{- \epsilon t} \Delta^{-i t}\dot{h} |\Psi\rangle,
\end{equation}
where $\widehat{a}$ is a possibly-unbounded modular zero mode affiliated with $\bar{\alg}.$

\subsection{Bulk reconstruction for full-rank codes}
\label{sec:full-rank-code-flow}

As in the previous section, let $\cH$ be a Hilbert space which contains two complementary subsystems represented by the algebras $\alg$ and $\bar{\alg}.$
Suppose that $\cH_{\text{code}} \subseteq \cH$ is a finite-dimensional subspace which is protected against the erasure of $\alg.$
This means that any operator acting within $\cH_{\text{code}}$ can be represented by an operator in $\bar{\alg}$. Suppose we are given two such code subspaces $\cH_{\text{code}}(0)$ and $\cH_{\text{code}}(1)$ with their respective operator reconstructions. We say that these two codes are smoothly connected if there exists a one-parameter family of code subspaces $\cH_{\text{code}}(\lambda)$ which interpolate between them that are all protected against erasure of $\alg$. Our goal is to relate the operator reconstructions along such a one-parameter family of codes.

Suppose further that we are given a preferred orthonormal basis $\{\ket{\psi_j(\lambda)}\}_{j=1}^d$ for each member of the family of code subspaces.
In the holographic context, this would be the case, for example, if our family of code subspaces were to represent a $d$-level quantum system located at a $\lambda$-dependent point in spacetime.
Fix a reference system $\cH_R$ of the same dimension as the code subspaces $\cH_{\text{code}}(\lambda),$ pick an orthonormal basis $\ket{j}$ for $\cH_R,$ and consider the one-parameter family of states
\begin{equation}
	|\Psi_{\lambda}\rangle
		= \frac{1}{\sqrt{d}} \sum_{j=1}^{d} |\psi_j(\lambda)\rangle \otimes |j\rangle.
\end{equation}
As explained in the introduction (eq. \ref{eq:transpose-trick-equation}), an operator $O_{\bar{A}}$ in $\bar{\alg}$ accurately reconstructs the operator 
$$\phi=\sum_{j,k} \phi_{jk} | \psi_j(\lambda) \rangle \langle \psi_k(\lambda) |$$  
on the code subspace if and only if it satisfies the equation
\begin{equation} \label{eq:recon}
	O_{\bar{A}} |\Psi_{\lambda} \rangle
		= \phi^T | \Psi_{\lambda} \rangle,
\end{equation}
where $\phi^T$ acts on $\cH_R$ and is given by
\beq 
\phi^T=\sum_{j, k} \phi_{kj} |j\rangle \langle k|_R.
\eeq 
We will fix a $\lambda$-independent set of matrix elements $\phi_{jk}$ for the code subspace operator we wish to reconstruct, but because the code subspace itself depends on $\lambda$, the reconstructed operator $O_{\bar{A}}$ will necessarily depend on $\lambda$. We can differentiate equation \eqref{eq:recon} to obtain a flow equation for $O_{\bar{A}}$:
\begin{equation}
	\dot{O}_{\bar{A}} |\Psi \rangle
	= \phi^T | \dot{\Psi} \rangle - O_{\bar{A}} |\dot{\Psi}\rangle.
\end{equation}
where we have suppressed the dependence on $\lambda$ to keep the notation simple.

If the code subspaces are such that the maximally entangled states $|\Psi_{\lambda}\rangle$ are cyclic and separating for $\bar{\alg},$ then we can go even further.
Plugging in the flow equation \eqref{eq:state-flow-with-zero-mode} from the previous subsection, and using the identity $(\phi^T - O_{\bar{A}}) \ket{\Psi} = 0,$ we obtain the following equation for the reconstructed operator
\begin{equation}
		\dot{O}_{\bar{A}} |\Psi\rangle = [ (\phi_T- O_{\bar{A}}),  \widehat{a}] |\Psi\rangle + i \int_0^{\infty} d t\, e^{- \epsilon t} [(\phi^T - O_{\bar{A}}), \Delta^{-i t}\dot{h} \Delta^{it}] |\Psi\rangle.
\end{equation}
Since $\widehat{a}$ is affiliated with $\bar{\alg}$ and $\phi^T$ acts on $\cH_R,$ we have $[\phi_T, \widehat{a}] = 0.$ In addition, the decoupling principle implies that modular flow acts trivially on the $R$ factor, so $\dot{h}$ commutes with operators in $R$. This then implies that 
$$[\phi^T, \Delta^{-it} \dot{h} \Delta^{it}] = 0.$$
This gives the simplified equation
\begin{equation} \label{eq:state-reconstruction-flow-penultimate}
	\dot{O}_{\bar{A}} |\Psi\rangle = [\widehat{a}, O_{\bar{A}}] |\Psi\rangle + i \int_0^{\infty} d t\, e^{- \epsilon t} [\Delta^{-i t}\dot{h} \Delta^{it}, O_{\bar{A}}] |\Psi\rangle.
\end{equation}
A further simplification can be obtained by observing that every code subspace operator $O_{\bar{A}}$ is fixed by modular flow.
This follows from the equation $O_{\bar{A}} \ket{\Psi} = \phi^T \ket{\Psi}$ and the fact that modular flow is trivial on the $R$ system, since we have
\begin{equation}
	\Delta^{-it} O_{\bar{A}} \Delta^{it} \ket{\Psi}
		= \Delta^{-it} \phi^T \Delta^{it} \ket{\Psi}
		= \phi^T \ket{\Psi}
		= O_{\bar{A}} \ket{\Psi}.
\end{equation}
Because $\ket{\Psi}$ is separating for $\bar{\alg},$ this implies the operator equation $\Delta^{-it} O_{\bar{A}} \Delta^{it} = O_{\bar{A}},$ which can be used to rewrite equation \eqref{eq:state-reconstruction-flow-penultimate} as
\begin{equation}
	\dot{O}_{\bar{A}} |\Psi\rangle = [\widehat{a}, O_{\bar{A}}] |\Psi\rangle + i \int_0^{\infty} d t\, e^{- \epsilon t} \Delta^{-i t} [\dot{h}, O_{\bar{A}}] \Delta^{it} |\Psi\rangle.
\end{equation}
Finally, since $\ket{\Psi}$ is separating for $\bar{\alg},$ and $\dot{O}_{\bar{A}}$ is in $\bar{\alg},$ the above vector equation implies the operator equation
\begin{equation} \label{eq:operator-flow-sec-2}
	\dot{O}_{\bar{A}} = [\widehat{a}, O_{\bar{A}}] + i \int_0^{\infty} d t\, e^{- \epsilon t} \Delta^{-i t} [\dot{h}, O_{\bar{A}}] \Delta^{it}.
\end{equation}

Equation \eqref{eq:operator-flow-sec-2} is a differential equation for the reconstructed operator $O_{\bar{A}}$ and is one of our main results. 
If $O_{\bar{A}}$ is known on the code subspace $\cH_{\text{code}}(0),$ then this equation can be integrated to give a reconstruction formula for $O_{\bar{A}}$ on the code subspace $\cH_{\text{code}}(1).$
In section \ref{sec:full-rank-codes-holography}, we will describe a setting in holography in which the code subspaces have cyclic-separating maximally entangled states, and thus for which equation \eqref{eq:operator-flow-sec-2} can be used for relational bulk reconstruction.

One comment is in order: in order to integrate this equation, it is necessary to know the modular operator $\Delta$ for each cyclic-separating state $|\Psi_{\lambda}\rangle.$
It is also necessary to know the operator $\widehat{a},$ which is related to the the projection of the vector $|\dot{\Psi}\rangle$ onto the zero-eigenvalue eigenspace of $h.$
It would be very convenient to be able to drop the term $[\widehat{a}, O_{\bar{A}}],$ since equation \eqref{eq:operator-flow-sec-2} would then be expressed solely in terms of modular flow.
We cannot simply accomplish this by assuming that there are no modular zero modes, because as discussed above, every reconstruction of a code subspace operator satisfying equation \eqref{eq:recon} is a zero mode for modular flow.
Instead, being able to neglect the first term in equation \eqref{eq:operator-flow-sec-2} is a condition on the kinds of flows $|\Psi_{\lambda}\rangle$ that we consider.
Since $\widehat{a} \ket{\Psi}$ is the projection of $|\dot{\Psi}\rangle$ onto the zero-mode eigenspace, we may neglect the first term in equation \eqref{eq:operator-flow-sec-2} if $|\dot{\Psi}\rangle$ has no support in the space of modular zero modes, or more generally if its projection to the space of modular zero modes commutes with all reconstructed operators. 

We claim that this restriction is actually quite natural.
One way that the first term in equation \eqref{eq:operator-flow-sec-2} could fail to vanish would be if the flow $\cH_{\text{code}}(\lambda)$ was produced by a unitary operator acting within the code subspace.
But the case of interest in quantum error correction is when we consider flows that involve no internal rotations within the code subspace; such internal rotations are simply relabelings of the code subspace basis, and do not have physical meaning.
With this in mind, suppose we consider only flows such that the deformation of the code subspace is orthogonal to the code subspace, i.e., deformations satisfying
\begin{equation} \label{eq:orthogonality-condition}
	\langle \dot{\psi}_j | \psi_k \rangle = 0.
\end{equation}
In particular, this condition implies that $|\dot{\Psi}\rangle$ is orthogonal to the code subspace.
Because the second term in equation \eqref{eq:state-flow-with-zero-mode} is already orthogonal to the code subspace, this implies that $\widehat{a} \ket{\Psi}$ is itself orthogonal to the code subspace.

If we are in a situation where the \textit{only} zero modes of modular flow are reconstructions of code subspace operators satisfying equation \eqref{eq:recon}, then the above restriction suffices to impose $\widehat{a} = 0.$
It is unclear to us whether this is true in general.
A state for which the only modular zero mode is the identity is called \textit{ergodic}, and while there appear to be no general theorems about when modular flow in quantum field theory is ergodic, ergodicity in modular flow is intimately connected to quantum chaos \cite{Gesteau:chaos, Ouseph:chaos}.
We might therefore expect that in holographic applications, the individual states $\ket{\psi_j}$ have no nontrivial zero modes, and the only nontrivial zero modes introduced in passing to the maximally entangled state $|\Psi\rangle$ are those in the code subspace.
This point certainly requires further investigation, but we believe it plausible that in applications to bulk reconstruction, the first term in equation \eqref{eq:operator-flow-sec-2} is zero when the orthogonality condition \eqref{eq:orthogonality-condition} is satisfied.

\subsection{Connection to the Connes cocycle flow}
\label{sec:connes-cocycle-connection}

We now explore an apparent connection between the flow equation \eqref{eq:operator-flow-sec-2} and the infinite-time limit of Connes cocycle flow.
This limit has previously appeared in \cite{Ceyhan:2018zfg} in a proof of the quantum null energy condition, in \cite{Levine:2020upy} in the setting of perturbative causal wedge expansion, and in \cite{Lashkari:2019ixo} in the context of weak sewing of states in quantum field theory.

Given commutant algebras $\alg, \bar{\alg},$ and two cyclic-separating state $\ket{\Phi_1}, \ket{\Phi_2}$, the Connes cocycle is a family of unitary operators in $\bar{\alg}$ that interpolates between the modular flows for the two states.
A detailed treatment of the Connes cocycle can be found in \cite[Chapter VIII.3]{Takesaki:volII}; we simply list the relevant properties of the cocycle, though note that we use different sign conventions than those chosen in \cite{Takesaki:volII}. 
The Connes cocycle is denoted $\bar{u}_{\Phi_2|\Phi_1}(t),$ and for $\bar{a} \in \bar{\alg},$ it satisfies
\begin{equation} \label{eq:cocycle-interpolation-identity}
	\Delta_{\Phi_2}^{-it} \bar{a} \Delta_{\Phi_2}^{it}
		= \bar{u}^{\dagger}_{\Phi_2|\Phi_1}(t) \left( \Delta_{\Phi_1}^{-it} \bar{a}\, \Delta_{\Phi_1}^{it} \right) \bar{u}_{\Phi_2|\Phi_1}(t).
\end{equation}
The dependence of $\bar{u}_{\Phi_2|\Phi_1}(t)$ on $t$ is not group-like, but rather satisfies the cocycle identity
\begin{equation} \label{eq:cocycle-multiplication-identity}
	\bar{u}_{\Phi_2|\Phi_1}(s + t)
		= \bar{u}_{\Phi_2|\Phi_1}(s) \left(\Delta_{\Phi_2}^{-is} \bar{u}_{\Phi_2|\Phi_1}(t) \Delta_{\Phi_2}^{is} \right),
\end{equation}
with $\bar{u}_{\Phi_2|\Phi_1}(0)$ equal to the identity.
The cocycle also satisfies a chain rule:
\begin{equation}
	\bar{u}_{\Phi_3|\Phi_1}(t)
	= \bar{u}_{\Phi_2|\Phi_1}(t) \bar{u}_{\Phi_3|\Phi_2}(t),
\end{equation}
and an adjoint identity:
\begin{equation}
	\bar{u}^{\dagger}_{\Phi_2|\Phi_1}(t)
		= \bar{u}_{\Phi_1|\Phi_2}(t).
\end{equation}

Suppose we now consider a one-parameter family of states $\ket{\Psi_{\lambda}},$ and employ the notation $\bar{u}_{\lambda} = \bar{u}_{\Psi_{\lambda}|\Psi_0}.$
Using equation \eqref{eq:cocycle-interpolation-identity}, we obtain the identity
\begin{equation}
    \bar{u}_{\lambda}^{\dagger}(s)\, \bar{a}\, \bar{u}_{\lambda}(s)
        = \Delta_{\lambda}^{-is} \Delta_0^{is} \bar{a} \Delta_0^{-is} \Delta_{\lambda}^{is}.
\end{equation}
Differentiating with respect to $\lambda$ and using $$\frac{\partial}{\partial \lambda} \Delta_{\lambda}^{is} = \Delta_{\lambda}^{is} \int_0^s dt\, \Delta_{\lambda}^{-i t} (- i \dot{h}_{\lambda}) \Delta_{\lambda}^{i t}$$ gives
\begin{equation} \label{eq:cocycle-operator-derivative} 
	\frac{\partial}{\partial \lambda} \left(\bar{u}_{\lambda}^{\dagger}(s)\, \bar{a}\, \bar{u}_{\lambda}(s)\right)
		= i \int_0^s dt\, [\Delta_{\lambda}^{-it} \dot{h}_{\lambda} \Delta^{i t}_{\lambda}, \bar{u}_{\lambda}^{\dagger}(s)\, \bar{a}\, \bar{u}_{\lambda}(s)].
\end{equation}
This is remarkably similar to our flow equation for reconstructed operators, which, in the setting where the zero-mode contribution can be neglected, is given by
\begin{equation} \label{eq:operator-flow-2-3}
	\frac{\partial}{\partial \lambda} O_{\bar{A}}
		= i \int_0^{\infty} dt\, e^{- \epsilon t} [\Delta_{\lambda}^{-it} \dot{h} \Delta_{\lambda}^{it}, O_{\bar{A}}].
\end{equation}
In particular, the equations seem to agree if we set $O_{\bar{A}}(\lambda) = \bar{u}_{\lambda}^{\dagger}(s) O_{\bar{A}}(0) \bar{u}_{\lambda}(s)$, ignore the regulating factor $e^{-\epsilon t}$, and take $s \to \infty$.

Naively, the above observations seem to suggest that if a pair of maximally entangled code subspace states $\ket{\Psi_0}, \ket{\Psi_1}$ is connected by a flow that does not generate zero modes, then a reconstructed operator $O_{\bar{A}}(1)$ can be expressed in terms of the reconstructed operator $O_{\bar{A}}(0)$ by taking the infinite-time limit of the Connes cocycle flow connecting $\ket{\Psi_0}$ to $\ket{\Psi_1}.$

There are two important caveats to this conclusion.
The first is that the formal manipulations above required neglecting the regulator term $e^{-\epsilon t}$, or equivalently, dropping the ``contour at infinity'', which may be important in general.
The second is that equation \eqref{eq:operator-flow-2-3} is a differential equation for the reconstructed operator $O_{\bar{A}}(\lambda)$, and even if the limit $s \to \infty$ can be taken at any fixed $\lambda$ to match the equations \eqref{eq:cocycle-operator-derivative} and \eqref{eq:operator-flow-2-3}, there is no guarantee that this limit commutes with the operation of integrating the differential equation to pass from $O_{\bar{A}}(0)$ to $O_{\bar{A}}(1).$
In other words, even if the error between equations \eqref{eq:cocycle-operator-derivative} and \eqref{eq:operator-flow-2-3} can be made arbitrarily small at fixed $\lambda$ by taking $s$ large, there is no guarantee that the error can be made \textit{uniformly} small in $\lambda,$ which would be necessary to prevent arbitrarily small errors at each $\lambda$ from accumulating to a nontrivial error in the integrated equation for $O_{\bar{A}}(1).$
Nevertheless, we find the similarity between equations \eqref{eq:cocycle-operator-derivative} and \eqref{eq:operator-flow-2-3}, together with the connection to the infinite-time cocycle flow used for relational reconstruction in \cite{Levine:2020upy}, extremely suggestive.

In our attempts to make this connection more rigorous, we have obtained some partial results that we state here in the interest of inspiring future developments.
These results also suggest a path to deriving the flow equation \eqref{eq:operator-flow-2-3} that does not require careless manipulations of unbounded operators.

The starting point of the more rigorous approach is the equation
\begin{equation} \label{eq:both-cocycles}
	\Delta_{\Phi_2}^{-it}
	= \bar{u}_{\Phi_2|\Phi_1}^{\dagger}(t) \Delta_{\Phi_1}^{-it} u_{\Phi_2|\Phi_1}(t),
\end{equation}
where we have introduced the symbol $u$ for the Connes cocycle with respect to the algebra $\alg.$
This equation holds exactly when $\alg$ and $\bar{\alg}$ are factors, meaning that the only operators in $\alg \cap \bar{\alg}$ are multiples of the identity; otherwise, equation \eqref{eq:both-cocycles} may be modified by terms involving operators in $\alg \cap \bar{\alg}$.
We will not prove equation \eqref{eq:both-cocycles} here, but the basic idea is to check that the right-hand side satisfies the KMS condition with respect to the state $\ket{\Phi_2}$, which implies the equality by the uniqueness of modular flow (see e.g. \cite[theorem 16]{Sorce:2023gio}).

For a one-parameter family of states $|\Psi_{\lambda}\rangle,$ equation \eqref{eq:both-cocycles} can be differentiated --- assuming the $\lambda$-derivative of the cocycle exists in some appropriate sense --- to obtain
\begin{equation}
	\partial_{\lambda} \Delta_{\lambda}^{-it}
	= \partial_{\lambda} \bar{u}_{\lambda}^{\dagger}(t) \Delta_{0}^{-it} u_{\lambda}(t)
	+ \bar{u}_{\lambda}^{\dagger}(t) \Delta_0^{-it} \partial_{\lambda} u_{\lambda}(t).
\end{equation}
Applying this to the derivative of the equation $\Delta_{\lambda}^{-it} |\Psi_{\lambda}\rangle = |\Psi_{\lambda}\rangle,$ suppressing the $\lambda$ subscripts, and using an overdot to denote a $\lambda$ derivative, we obtain
\begin{equation} \label{eq:rigorous-flow}
	| \dot{\Psi} \rangle
		= \left( \dot{\bar{u}}^{\dagger}(t) \bar{u}(t) + \Delta^{-it} u^{\dagger}(t) \dot{u}(t) \right)|\Psi \rangle
	+ \Delta^{-it} |\dot{\Psi} \rangle.
\end{equation}
This is a version of the state-flow equation which is completely well founded; our previous equation \eqref{eq:state-flow-intro} can be roughly thought of as the analytic continuation of this equation to $t=i/2.$ 

Now let us return to the basic equations for a reconstructed operator,
\begin{align}
	0
	& = (\phi^T - O_{\bar{A}}) |\Psi \rangle, \\
	\dot{O}_{\bar{A}} |\Psi \rangle
	& = (\phi^T - O_{\bar{A}}) |\dot{\Psi} \rangle.
\end{align}
Plugging in equation \eqref{eq:rigorous-flow} and performing the same kinds of manipulations as in section \ref{sec:full-rank-code-flow}, we obtain
\begin{equation} \label{eq:rigorous-operator-flow}
	\dot{O}_{\bar{A}} |\Psi \rangle
		= [\dot{\bar{u}}^{\dagger}(t) \bar{u}(t), O_{\bar{A}}] |\Psi \rangle
		+ (\phi^T - O_{\bar{A}})  \Delta^{-it} |\dot{\Psi} \rangle.
\end{equation}

Now we ask: if we define $O_{\bar{A}}$ as
\begin{equation} \label{eq:O-cocycle-definition}
	O_{\bar{A}}(\lambda)
		= \bar{u}_{\lambda}^{\dagger}(t) O_{\bar{A}}(0) \bar{u}_{\lambda}(t),
\end{equation}
then how close does it come to satisfying equation \eqref{eq:rigorous-operator-flow}?
Differentiating equation \eqref{eq:O-cocycle-definition}, we obtain
\begin{equation}
	\dot{O}_{\bar{A}}
		= \dot{\bar{u}}^{\dagger}(t) \bar{u}(t) O_{\bar{A}}
			+ O_{\bar{A}} \bar{u}^{\dagger}(t)\dot{\bar{u}}(t).
\end{equation}
Using unitarity of the cocycle flow, it is easy to show that $\dot{\bar{u}}^{\dagger}(t) \bar{u}(t)$ is anti-Hermitian, so this may be expressed as
\begin{equation}
	\dot{O}_{\bar{A}}
	= [\dot{\bar{u}}^{\dagger}(t) \bar{u}(t), O_{\bar{A}}].
\end{equation}
It follows that equation \eqref{eq:O-cocycle-definition} exactly reproduces equation \eqref{eq:rigorous-operator-flow} up to the term $(\phi^T - O_{\bar{A}}) \Delta^{-it} |\dot{\Psi} \rangle.$
If we were able to show that this term becomes uniformly small in $\lambda$ as $t$ is taken large, then we would immediately be able to conclude that the reconstructed operator is the infinite-cocycle-time limit of $O_{\bar{A}}(1).$
We are uncertain as to exactly what conditions are necessary for this to occur, but if we approximate $|\dot{\Psi}\rangle$ in terms of an operator in $\bar{A}$ as $|\dot{\Psi}\rangle \approx \bar{a}|\Psi\rangle,$ then the condition can be stated as
\begin{equation} \label{eq:uniform-decay-condition}
	[O_{\bar{A}}, \Delta^{-it} \bar{a}\Delta^{-it}] |\Psi \rangle \to 0 \quad \text{uniformly in $\lambda$ in the limit $t \to \infty$}.
\end{equation}
For ergodic modular flows, the infinite-modular-time limit of any operator tends to decay in correlation functions --- see e.g. \cite{Gesteau:chaos, Ouseph:chaos}.
So, by the same arguments made in the previous subsection in order to neglect the zero-mode term, it seems plausible that condition \eqref{eq:uniform-decay-condition} will be satisfied in reasonable holographic setups.
Much more must be done to establish this claim with certainty.

\section{Full-rank codes in holography}
\label{sec:full-rank-codes-holography}

In the previous section, we derived a flow equation satisfied by reconstructed operators for certain kinds of error-correcting codes.
The fundamental assumptions were (i) that the code was exactly protected against erasures of a region $A$, and (ii) that the maximally entangled state with a reference, $|\Psi\rangle$, was cyclic and separating (i.e., full rank) with respect to the bipartition into $\bar{A}$ and $A \cup \text{reference}$.
We will now argue that this setup can be applied to physically interesting problems in holographic quantum gravity.

Because we are assuming exact error correction, we must work in the infinite-N ($G_N \to 0$) limit of the holographic theory.
In this setting, one considers some single-trace operators which admit a good large-$N$ limit, and one constructs a Hilbert space by acting with these operators on a semi-classical background state, such as, for instance, the vacuum state. The Hilbert space generated this way is isomorphic to the Hilbert space of bulk quantum fields in a particular, fixed background geometry.
Traditionally, the task of bulk reconstruction is to start with a quantum field operator in the bulk region $\bar{a}$, and to produce a formula for the equivalent operator in the boundary region $\bar{A}.$
It is generally assumed that a typical state of the bulk quantum fields will be cyclic and separating for $\bar{a}.$

This assumption, however, prevents us from building nontrivial quantum error-correcting codes of the kind we considered in the previous section.
Suppose we have two code subspace states $|\psi_1 \rangle, |\psi_2\rangle,$ and denote the corresponding maximally entangled state by
\begin{equation}
	|\Psi\rangle
		= \frac{|\psi_1\rangle |1\rangle + |\psi_2\rangle |2\rangle}{\sqrt{2}}.
\end{equation}
If this code is correctable against the erasure of $A$, then the decoupling principle states that for any operator $a$ acting on $A$, we must have
\begin{equation}
	\langle \Psi | (a \otimes |j\rangle \langle k |) |\Psi \rangle
		= \langle \Psi| a | \Psi \rangle \langle \Psi | j \rangle \langle k | \Psi \rangle,
\end{equation}
which gives
\begin{equation}
	\langle \psi_j | a | \psi_k \rangle
		= \delta_{jk} \frac{\langle \psi_1 | a | \psi_1 \rangle + \langle \psi_2 | a | \psi_2 \rangle}{2}.
\end{equation}
In particular, this implies $\langle \psi_1 | a | \psi_2 \rangle = 0$ for any operator $a$ acting on $A$.
If $|\psi_2 \rangle$ is cyclic for $A$, this implies that $|\psi_1 \rangle$ vanishes; if $|\psi_1 \rangle$ is cyclic for $A$, this implies that $|\psi_2 \rangle$ vanishes.
In either case, we have demonstrated that it is impossible to build an exact, isometric quantum error correcting code out of cyclic and separating states.

The above arguments do not necessarily preclude the setup considered in section \ref{sec:main-section}, which only required that the maximally entangled state with a reference be cyclic and separating with respect to the division into $\bar{A}$ and $A \cup \text{reference}$. 
It is possible that there are physically interesting scenarios in AdS/CFT where none of the code subspace states are cyclic and separating, but the maximally entangled state is.
However, we argue now that there is a natural setting for bulk reconstruction in which the maximally entangled state is guaranteed to be cyclic and separating.

Suppose we start with a holographic state $|\psi \rangle_{A\bar{A}}$ which is cyclic and separating for the division of the system into $A$ and $\bar{A}$.
We will build up a code subspace by introducing an auxiliary qudit $\cH_{Q}$ and coupling it to $|\psi \rangle_{A\bar{A}}$. The way we do this is as follows: we consider some bulk mode well-localized within the entanglement wedge $\bar{a}$ of $\bar{A}$, and we couple the auxiliary qudit to this bulk mode using a unitary operator in $\bar{a}Q.$
That is, we consider a code subspace with orthonormal basis given by
\begin{equation}
	|\psi_j\rangle_{A\bar{A} Q} = U_{\bar{a}Q} (\ket{\psi}_{A \bar{A}} |j\rangle_{Q}).
\end{equation}
We will take the $\{\psi_j\}$ as our code subspace states in the enlarged Hilbert space $\cH_{\text{CFT}}\otimes \cH_Q$. The maximally entangled state with respect to a reference $R$ is simply
\begin{equation}
	|\Psi\rangle
		= U_{\bar{a} Q}( \ket{\psi}_{A \bar{A}} \otimes |\text{MAX}\rangle_{QR}).
\end{equation}
Since $|\psi\rangle$ is cyclic and separating for the bi-partition $A:\bar{A},$ and $|\text{MAX}\rangle$ is cyclic and separating for the bi-partition $Q:R,$ the state $|\Psi\rangle$ is cyclic and separating for the bi-partition $AR:\bar{A}Q$.

 Given a cyclic-separating holographic state $|\psi\rangle$ and a coupling unitary $U_{\bar{a} Q},$ we can construct code subspaces of this type. For such code subspaces, we conclude that it is possible to reconstruct arbitrary code subspace operators on $\bar{A}Q$ using the flow equations derived in section \ref{sec:main-section}.
While these reconstructed operators do not live in the holographic CFT, they can be used to \textit{measure} operators in the holographic CFT to arbitrary precision.
To be more precise, suppose $\bar{a}$ is a bulk observable with spectral decomposition $\bar{a} = \sum_{n} \alpha_n |\phi_n\rangle \langle \phi_n|,$ and $|\psi\rangle$ is a state written in the $\bar{a}$-eigenbasis as
\begin{equation}
    |\psi\rangle 
        = \sum_n c_n |\phi_n\rangle.
\end{equation}
Measuring the operator $\bar{a}$ means sampling the distribution of coefficients $|c_n|^2.$
To approximate this, we can couple $d$ eigenvalues of $\bar{a}$ to an auxiliary qudit in state $|\chi\rangle$ by acting on $|\psi\rangle \otimes |\chi\rangle$ as
\begin{equation}
    e^{-i t \sum_{j=1}^{d} |\phi_{n_j} \rangle \langle \phi_{n_j} | \otimes \sigma_{n_j}} |\psi\rangle \otimes | \chi \rangle
        = \sum_{n} c_n |\phi_n\rangle e^{- i t \sigma_{n}} |\chi\rangle.
\end{equation}
where the $\sigma_{n_j}$ are Hermitian operators acting on the auxiliary qudit.
The reduced density matrix of this state is
\begin{equation}
    \sum_{n} |c_n|^2 |e^{- i t \sigma_n} \chi \rangle \langle e^{- i t \sigma_n} \chi|.
\end{equation}
By increasing the dimension $d$ of the qudit and tuning the operators $\sigma_n$ and the time $t,$ the states $|e^{- i t \sigma_n} \chi \rangle$ can be made approximately pairwise orthogonal, and the distribution of coefficients $|c_n|^2$ can be sampled by measuring an orthonormal basis of the qudit.
This is exactly the model of quantum measurement advocated by von Neumann, where an arbitrary quantum system is measured by coupling it to an auxiliary detector whose energy levels can be observed.

We note that the measurement-operator paradigm we have described here bears a strong resemblance to the framework for bulk reconstruction developed in \cite{Lamprou:probes1, Lamprou:probes2, Lamprou:probes3}, in which features of a large black hole are measured by coupling to a probe black hole.
As the formulas in those papers also involve modular flow, it would be interesting to make a more explicit connection between our work and theirs.

\section{Bulk reconstruction from modular reflections}
\label{sec:modular-reflections}

In this section, we give an alternative derivation of some of the essential formulas from section \ref{sec:main-section} in the non-algebraic, density-matrix formalism.
One of our motivations is that while the algebraic techniques of section \ref{sec:main-section} rely strongly on features of infinite $N$ and exact quantum error correction, the techniques presented here may generalize more easily. Along the way, we will also introduce a way to think about bulk reconstruction in terms of canonical purifications, following the exposition in \cite{Harlow:2016vwg}. 

We begin in section \ref{sec:reflection-decoupling} by defining the modular reflection operator, showing that it can be used as a universal recovery map for erasure codes, and explaining a connection to the decoupling principle.
In section \ref{sec:reflection-operator-flow}, we derive a flow equation for the reflection operator (discussed previously in \cite{Parrikar:quantum-extremal-shock}) and show that it reproduces our main equation \eqref{eq:operator-flow-intro}.

\subsection{The reflection operator and decoupling}	
\label{sec:reflection-decoupling}

Consider the Euclidean path integral in a general quantum field theory on the manifold $I \times \Sigma$, where $\Sigma$ is a compact spatial manifold and $I$ is a segment of Euclidean time $\beta/2$.
Let $\cH$ be the Hilbert space of the theory on $\Sigma$.
This path integral can be thought of as the map
\begin{equation}
\Psi_0: \cH \to \cH,\;\;\; \Psi_0 = e^{-\frac{\beta}{2}H}.
\end{equation}
Equivalently, we can think of $\Psi_0$ as a state in the doubled Hilbert space $\cH \otimes \cH^{\st}$, where $\cH^{\st}$ is the dual Hilbert space to $\cH$.
$|\Psi_0\rangle$ is called the thermofield double state, and is often written as
\begin{equation}
|\Psi_0\rangle \sim \sum_{n}e^{-\frac{\beta}{2}E_n} |n\rangle\otimes |n^{\st}\rangle,
\end{equation}
where $E_n$ are the energy eigenvalues of the Hamiltonian, $|n\rangle$ are the energy eigenstates, and $|n^{\st}\rangle$ is the dual vector corresponding to the eigenstate $\ket{n}$, i.e., $|n^{\st}\rangle = \bra{n}$.
The reduced density matrix of the state $|\Psi_0\rangle$ on the first factor is the thermal ensemble:
\begin{equation}
\rho_0 \sim \sum_{n}e^{- \beta E_n} |n\rangle\langle n|.
\end{equation}
From this point of view, the thermofield double can be thought of as a purification of $\rho_0$.

Given a general density matrix $\rho$ on $\cH$, a purification of $\rho$ is a pure state $\ket{\Psi}$ on a larger system $\cH \otimes \cH'$ satisfying
\begin{equation}
	\Tr_{\cH'} \ket{\Psi}\bra{\Psi} = \rho.
\end{equation}
While there are infinitely many purifications for any density matrix $\rho$, there always exists a \textit{canonical purification} given by the operator $\sqrt{\rho}$, treated as a vector in $\cH \otimes \cH^\st$.
The thermofield double state $\ket{\Psi_0}$ is the canonical purification for the thermal ensemble $\rho_0$.

Let us now consider a general state $|\Psi\rangle$ of a bipartite Hilbert space $\cH_A \otimes \cH_{\bar A}$.
The reduced density matrix on the first factor is obtained by tracing out the second factor:
\begin{equation}
	\rho_{A}(\Psi)
		= \mathrm{Tr}_{\cH_{\bar A}} |\Psi\rangle \langle \Psi|.
\end{equation}
We can also define the canonical purification of $\Psi$ with respect to $A$ as the vector  $\Psi^{\st}= \sqrt{\rho_{A}^{\Psi}} \in \cH_A \otimes \cH^\st_A$, where again $\cH_A^\st$ is the dual Hilbert space corresponding to $\cH_A$. In the case of the thermofield double state, the canonical purification and the original state were the same.
In general, however, $\Psi$ and  $\Psi^{\st}$ will be different purifications of the same density matrix $\rho_{A}(\Psi)$.
This implies that there exists an operator $\cR$ satisfying
\begin{equation} \label{MROdef}
|\Psi\rangle = \cU\,|\Psi^\st\rangle, 
\end{equation}
where $\cR$ acts only on the second factor, i.e., 
\begin{equation}
\cU: \cH_{A}^{\st} \to \cH_{\bar A}.
\end{equation}
If $\Psi$ is full-rank with respect to $A$ (i.e., if $\rho_A(\Psi)$ has no zero eigenvalues), then the action of $\cU$ is uniquely determined on all of $\cH^\st_A$.
It is also easily seen that $\cU$ is an isometry, i.e., $\cU^{\dagger}\cU = 1$.
The operator $\cR$ will be called the \emph{modular reflection operator} of the state $\Psi$ with respect to $A$.
(See Appendix \ref{sec:appB} for an algebraic approach to defining the operator $\cU$.)

Consider now a finite-dimensional subspace $\cH_{\text{code}} \subseteq \cH_{A} \otimes \cH_{\bar{A}},$ a reference system $R$, and a maximally entangled state
\begin{equation}
	|\Psi\rangle
		= \frac{1}{\sqrt{d}} \sum_{j=1}^{d} |\psi_j\rangle_{A\bar{A}}\otimes |j\rangle_R,
\end{equation}
where $|\psi_j\rangle$ is an orthonormal basis for $\cH_{\text{code}}$ and $|j\rangle$ is an orthonormal basis for $\cH_{R}.$
In the introduction, we mentioned the famous decoupling principle of quantum error correction, which states that $\cH_{\text{code}}$ is corrected against erasures of $A$ if and only if the maximally entangled state $|\Psi\rangle$ decouples between $A$ and $R$:
\begin{equation} \label{eq:correlator-decoupling}
	\langle \Psi | O_{A} \otimes O_R |\Psi\rangle
		= \langle \Psi | O_A |\Psi \rangle \langle \Psi | O_R |\Psi \rangle.
\end{equation}
We can now give a proof of this equivalence which directly involves the modular reflection operator of $|\Psi\rangle$ with respect to the subsystem $A \cup R,$ and which shows that the reflection operator can be used for quantum error correction. This argument is essentially a slight variant of the argument presented in \cite{Harlow:2016vwg}.

The direction ``protected against erasures of $A$ $\Rightarrow$ decoupling'' proceeds almost exactly as in \cite{Harlow:2016vwg}.
We say the code subspace $\cH_{\text{code}}$ is protected against erasures of $A$ if for any operator $\phi = \sum_{jk}\phi_{jk} |\psi_j\rangle \langle \psi_k |$ acting on the code subspace, there exists an operator $O_{\bar{A}}$ in $\bar{A}$ satisfying
\begin{align}
	O_{\bar{A}}\, P_{\text{code}}
		& = \phi\, P_{\text{code}}, \\
	O_{\bar{A}}^{\dagger}\, P_{\text{code}}
		& = \phi P_{\text{code}}.
\end{align}
In the above expressions, $P_{\text{code}}$ is the orthogonal projector onto the code subspace.
For any operators $O_{A}$ in $A$ and $O_{R}$ in $R$, because the maximally entangled state $|\Psi\rangle$ is fixed by $P_{\text{code}},$ we have
\begin{equation} \label{eq:P-code-insertion}
	\bra{\Psi} O_{A} O_{R} \ket{\Psi}
	= \bra{\Psi} (P_{\text{code}} O_{A} P_{\text{code}}) O_{R} \ket{\Psi}.
\end{equation}
The operator $P_{\text{code}} O_A P_{\text{code}}$ must be proportional to $P_{\text{code}},$ since otherwise there would exist a state $\ket{\chi} \in \cH_{\text{code}}$ and an operator $L_{\text{code}}$ acting on $\cH_{\text{code}}$ with
\begin{equation}
	0
	\neq \bra{\chi} [P_{\text{code}} O_A P_{\text{code}}, L_{\text{code}}] \ket{\chi}
	= \bra{\chi} [O_A, L_{\text{code}}] \ket{\chi},
\end{equation}
which contradicts the error-correction assumption that $L_{\text{code}}$ can be expressed as an operator acting on $\cH_{\bar{A}}.$
This observation gives the identity
\begin{equation}
	P_{\text{code}} O_A P_{\text{code}}
	= \bra{\Psi} O_A \ket{\Psi} P_{\text{code}},
\end{equation}
which combined with equation \eqref{eq:P-code-insertion} yields the desired equation \eqref{eq:correlator-decoupling}.

Conversely, we now suppose that the state $\ket{\Psi}$ decouples between $\cH_{A}$ and $\cH_{R}$ as in equation \eqref{eq:correlator-decoupling}.
This implies that the density matrix $\rho_{A,R}^{\Psi}$ factorizes as $\rho_{A}^{\Psi} \otimes \rho_{R}^{\Psi}.$
It follows that the canonical purification of $\rho^{\Psi}_{A, R}$ factorizes as well, and can be written in the form
\begin{equation}
	\ket{\Psi^{\st}}
	=  |\boldsymbol{\chi}\rangle_{A,A^\st} \otimes \left( \frac{1}{\sqrt{d}}\sum_{j} \ket{j}_{R} \otimes \ket{j^{\st}}_{R^{\st}} \right),
\end{equation}
where $\chi$ is the canonical purification of the $A$ factor. Let $\cR : \cH_{A^{\st}, R^{\st}} \to \cH_{\bar{A}}$ be the modular reflection operator mapping $\ket{\Psi^{\st}}$ to $\ket{\Psi}.$
Because $\cR$ does not act on $\cH_{R},$ the basis states for $\cH_{\text{code}}$ can be obtained via the formula
\begin{equation} \label{eq:mod-reflection-factorized}
	\ket{\psi_j}
	= \cR (|\boldsymbol{\chi}\rangle_{A, A^{\st}} \otimes \ket{j^{\st}}_{R^\st}).
\end{equation}
Consequently, if $\phi$ is the operator on $\cH_{\text{code}}$ with matrix elements
\begin{equation}
	\phi_{jk} 
	= \bra{\psi_j} \phi \ket{\psi_k},
\end{equation}
then the operator
\begin{equation} \label{eq:explicit-reflection-formula}
	O_{\bar{A}}
	= \cR \left(\sum_{j, k} \phi_{jk} \ket{j^{\st}}_{R^{\st}} \bra{k^{\st}}_{R^{\st}} \right) \cR^{\dagger}
\end{equation}
provides an explicit representation of $\phi$ in terms of an operator acting on $\bar{A}$. Thus, the modular reflection operator for the maximally entangled state on the code subspace gives an explicit reconstruction map on the code subspace. In the exact error correction setting, the reconstructed operators we get this way agree with the Petz recovery map.

Recall that in the context of holography, the operators acting within $\cH_{\text{code}}$ are thought of as bulk operators acting in a bulk region $\bar{a}.$ To be concrete, we can consider the code subspace to be the states of a bulk qudit well-localized within $\bar{a}$. In this situation, the Ryu-Takayanagi formula with quantum corrections implies that the mutual information between the reference system and $A$ vanishes:
\begin{eqnarray}
I_{\Psi}(A:R) &=& S_{\Psi}(R)+S_{\Psi}(A) - S_{\Psi}(A \cup R)\nonumber\\
&=& \log\,d + S_{\Psi}(A)-S_{\Psi}(\bar{A})\nonumber\\
&=& \log\,d + \frac{\text{Area}}{4G_N} - \left(\frac{\text{Area}}{4G_N} + \log\,d\right)\nonumber\\
&=& 0,
\end{eqnarray}
where in the third line we have used the quantum RT formula for the boundary subregions $A$ and $\bar{A}$, with the $\log d$ in the last term coming from the bulk entanglement corresponding to the maximally mixed bulk qudit contained in $\bar{a}$.
The vanishing of this mutual information implies that $|\Psi\rangle$ decouples between $A$ and $R$, which implies that the bulk qudit in $\bar{a}$ is protected against the erasure of the boundary system $A.$
It follows that \eqref{eq:explicit-reflection-formula} gives an explicit formula for the boundary representation of bulk operators in terms of the modular reflection operator $\cR$ for the maximally entangled state of the code subspace with a reference system. In appendix \ref{sec:app}, we give an alternative explanation for how the modular reflection operator accomplishes bulk reconstruction in holography, in terms of the replica trick.
While the reflection operator is only uniquely determined when $|\Psi\rangle$ is full-rank --- see again section \ref{sec:full-rank-codes-holography} for a discussion of this assumption --- in the case of a non-full-rank state $|\Psi\rangle,$ there is still a family of reflection operators that differ on the zero-Schmidt-eigenspace, any of which can be used as a reconstruction map.

We emphasize that any purification of $|\Psi\rangle$ could have been used to produce a reconstruction map using the above procedure.
There is nothing particularly special about the canonical purification except that it can be written using explicit formulas.

\subsection{A flow equation for the reflection operator}
\label{sec:reflection-operator-flow}

Suppose that, as in section \ref{sec:main-section}, we have two full-rank states $|\Psi_0\rangle$ and $|\Psi_1\rangle$ in the Hilbert space $\cH_A \otimes \cH_{\bar A}$.\footnote{Note that when we specialize to the case where $|\Psi\rangle$ is a maximally entangled state for a code, the system we are now calling $\cH_A$ will include both the erasable region previously called $A$ and the reference system $R.$ Similarly, if the code is constructed by bringing in an auxiliary qudit, then $\bar{A}$ will include the CFT region complementary to $A$ as well as the auxiliary qudit $Q$.}
As before, we will assume that these states are smoothly connected, meaning there exists a family $|\Psi_{\cl}\rangle$ of full-rank states interpolating between $|\Psi_0\rangle$ and $|\Psi_1\rangle$. We will also assume that the Schmidt spectrum of $|\Psi_{\cl}\rangle$ depends smoothly on $\lambda$ without encountering any level crossings.
In particular, degeneracies, if any, are not broken anywhere along the flow.

At any value of $\cl$, we can construct the reduced density matrices $\rho_A(\cl)$ and $\rho_{\bar A}(\cl)$ corresponding to the tensor factors $\cH_A$ and $\cH_{\bar A}$. Accordingly, we have a one-parameter family of one-sided modular Hamiltonians $K_A(\cl)$ and $K_{\bar A}(\lambda)$, where the one-sided modular Hamiltonian of a density matrix $\rho$ is defined as $K=-\log\,\rho$.
At any given value of $\cl$, we have a Schmidt decomposition for the state $|\Psi_{\cl}\rangle$:
\begin{equation} \label{PsiState}
	|\Psi_{\cl}\rangle
		= \sum_{n}\sum_j e^{-\frac{1}{2}E_n(\cl)}|\ch_{n,j}(\cl)\rangle\otimes |\tch_{n,j}(\cl)\rangle,
\end{equation}
where $n$ is a label for the modular eigenvalues, and $j$ is a label for degeneracies, the number of which can depend on $n$.

In terms of the one-sided modular Hamiltonians, the states $|\ch_{n, j}\rangle$ and $|\tch_{n, j}\rangle$ satisfy
\beq
K_A(\cl)|\ch_{n,j}(\cl)\rangle = E_n(\cl) |\ch_{n,j}(\cl)\rangle,
\eeq
\beq
K_{\bar{A}}(\cl)|\tch_{n,j}(\cl)\rangle = E_n(\cl) |\tch_{n,j}(\cl)\rangle.
\eeq
Using standard time-independent quantum mechanical perturbation theory, one can show that the modular eigenstates of the state $|\Psi_\cl\rangle$ satisfy the equation
\beq \label{eq:QM-pert-theory}
\frac{d}{d\lambda} |\ch_{n,j}\rangle  = i\,a |\ch_{n,j}\rangle+ \sum_{m\neq n}\sum_{k}\frac{\langle \ch_{m,k} | \frac{d}{d\lambda} K_A |\ch_{n,j}\rangle}{(E_n(\lambda)-E_m(\lambda))} |\ch_{m,k}\rangle.
\eeq
In the first term, $a$ is a Hermitian operator on $A$ that is block-diagonal in the modular energy basis.
Note that $a$ is a modular zero mode, as it commutes with the one-sided modular Hamiltonian $K_A$.
By using the identity
\beq 
\frac{1}{E_n - E_m} = \lim_{\epsilon \to 0^+} \pm \int_0^{\infty} idt\, e^{-\epsilon t \pm i(E_m-E_n)t},
\eeq 
we can rewrite equation \eqref{eq:QM-pert-theory} as
\beq
\frac{d}{d\lambda} |\ch_{n,j}\rangle
	= i\,a |\ch_{n,j}\rangle
	- \lim_{\epsilon \to 0^+} i \int_0^\infty e^{- \epsilon t} \sum_{m\neq n}\sum_{k} \Big|\ch_{m,k}\Big\rangle \Big \langle \ch_{m,k} \Big| e^{- i K_A t} \dot{K}_A e^{i K_A t} \Big|\ch_{n,j}\Big\rangle,
\eeq
where we have adopted the notation $\dot{K}_A = \frac{d K_A}{d \lambda}.$
By adding and subtracting an $m=n$ term for the sum, and taking the $\epsilon \to 0^+$ limit to be implicit, we may simplify this expression to
\beq
\frac{d}{d\lambda} |\ch_{n,j}\rangle  = i \left( a - \frac{1}{\epsilon} \dot{E}_n \right) |\ch_{n,j}\rangle - i \int_0^\infty e^{- \epsilon t} e^{- i K_A t} \dot{K}_A e^{i K_A t} |\ch_{n,j}\rangle.
\eeq
Going forward we will further simplify this expression by rewriting $a_{n;jk}^{\text{new}} = (a_{n;jk}^{\text{old}} -\frac{1}{\epsilon}\dot{E}_n \delta_{jk}),$ so that we may write
\beq \label{eq:chi-flow}
\frac{d}{d\lambda} |\ch_{n,j}\rangle  = i\, a |\ch_{n,j}\rangle - i \int_0^\infty e^{- \epsilon t} e^{-i K_A t} \dot{K}_A e^{i K_A t} |\ch_{n,j}\rangle.
\eeq
A similar formula is also true for the modular eigenstates of the $\bar{A}$ factor:
\beq \label{eq:tchi-flow}
\frac{d}{d\lambda} |\tch_{n,j}\rangle
=i \bar{a} |\tch_{n,j}\rangle + i \int_0^{\infty} dt\,e^{-\epsilon t} e^{i K_{\bar A} t} \dot{K}_{\bar A}  e^{-i K_{\bar A} t}|\tch_{n,j}\rangle,
\eeq
with $\dot{K}_{\bar A} = \frac{dK_{\bar{A}}}{d\lambda}$, and $ \bar{a}$ is a modular zero mode acting on the factor $\cH_{\bar{A}}.$

Taking the derivative of equation \eqref{PsiState} and plugging in these formulas gives
\begin{equation}
	|\dot{\Psi}\rangle
		= \widehat{a} |\Psi\rangle
			+ i \int_0^{\infty} dt\, e^{-\epsilon t}
			\left( e^{i K_{\bar{A}} t} \dot{K}_{\bar{A}} e^{-i K_{\bar{A}} t} - e^{-i K_A t} \dot{K}_{A} e^{i K_A t} \right) |\Psi\rangle,
\end{equation}
with
\begin{equation}
	\widehat{a}
		= - \frac{1}{2} \sum_{n, j} \dot{E}_{n} |\chi_{n,j}\rangle \langle \chi_{n, j}|
		+ i (a + \bar{a}). 
\end{equation}
Since $K_{\bar{A}}$ and $K_A$ commute, we may rewrite this in terms of the full modular Hamiltonian $h = K_{\bar{A}} - K_A$ as
\begin{equation}
	|\dot{\Psi}\rangle
	= \widehat{a} |\Psi\rangle
	+ i \int_0^{\infty} dt\, e^{-\epsilon t}
	e^{i h t} \dot{h} e^{-i h t} |\Psi\rangle,
\end{equation}
which matches the fundamental equation \eqref{eq:state-flow-with-zero-mode} once we use the identities $\Delta^{-it} = e^{i h t}$ and $h |\Psi\rangle = 0.$
The operator $\widehat{a}$ appearing in this equation is different than the one appearing in equation \eqref{eq:state-flow-with-zero-mode}, since there $\widehat{a}$ was a zero mode supported on $\bar{A}$, and here $\widehat{a}$ has support on both $\bar{A}$ and $A$.
But since $|\Psi\rangle$ is full rank, the action of any operator in $A$ on $|\Psi\rangle$ can be equivalently expressed as the action of a ``transposed'' operator in $\bar{A},$ and once this change is made the two formulas are exactly the same.

We could proceed as in section \ref{sec:full-rank-code-flow} to obtain equation \eqref{eq:operator-flow-intro} for the flow of reconstructed operators, but since we know from section \ref{sec:reflection-decoupling} that the reflection operator of an erasure code furnishes a reconstruction map, we will here derive equation \eqref{eq:operator-flow-intro} in terms of a flow equation for the reflection operator.
In terms of the Schmidt basis, the modular reflection operator $\cU_{\cl}$ is defined as the isometry
\begin{equation} \label{eq:MROFacdef}
	\cU
	= \sum_{n, j} |\tch_{n, j}\rangle \langle \ch_{n, j}^{\st}|,
\end{equation}
where $|\ch_{n, j}^{\st}\rangle \in \cH^{\st}_A$ is the dual vector corresponding to $|\chi_{n, j}\rangle$. 
To lighten the notation in equation \eqref{eq:chi-flow} and \eqref{eq:tchi-flow}, we will now write
\begin{align}
	\cA
	& = i \left(a - \int_0^{\infty} e^{-\epsilon t} e^{-i K_A t} \dot{K}_{A} e^{i K_A t}\right), \label{eq:cA-eq} \\
	\bar{\cA}
	& = i \left(\bar{a} + \int_0^{\infty} e^{-\epsilon t} e^{i K_{\bar{A}} t} \dot{K}_{\bar{A}} e^{-i K_{\bar{A}} t}\right),
\end{align}
so that the above equations become $|\dot{\chi}_{n, j}\rangle = \cA |\chi_{n, j}\rangle$ and $|\dot{\bar{\chi}}_{n, j} \rangle = \bar{\cA} |\tch_{n, j}\rangle.$
The derivative of the reflection operator is given by
\begin{equation}
	\dot{\cU}
		= \bar{\cA} \cU + \cU (\cA^{\st})^{\dagger},
\end{equation}
where $\cA^{\st}$ is given by a version of equation \eqref{eq:cA-eq} with all unstarred quantities given by their starred counterparts.\footnote{Equivalently, $\cA^{\st}$ is the conjugation of $\cA$ by the canonical antiunitary map taking kets to bras.}

In the setting of quantum erasure correction, where the system $A$ is enlarged to $AR$ and the state $|\Psi\rangle$ factorizes between $A$ and $R$, we know from section \ref{sec:reflection-decoupling} that a reconstructed operator $O_{\bar{A}}$ can be expressed as
\begin{equation}
	O_{\bar{A}}
		= \cU( \phi_{R^\st}\otimes \mathbb{1}_{A^\st}) \cU^{\dagger},
\end{equation}
where $\phi$ is an operator on $R^{\st}$ with fixed matrix elements.
Taking a $\lambda$ derivative of this equation gives
\begin{equation}
	\dot{O}_{\bar{A}}
	= \bar{\cA} O_{\bar{A}} + O_{\bar{A}} \bar{\cA}^{\dagger} + \cU (\cA^{\st})^{\dagger} \phi \cU^{\dagger} + \cU \phi \cA^{\st} \cU^{\dagger}.
\end{equation}
$|\Psi\rangle$ is maximally mixed on $R$, modular flow is trivial on $R^{\st}$, so all of the terms in $\cA^{\st}$ involving modular flow commute with $\phi.$
We also know that $a^{\st}$ is Hermitian by construction.
This leaves us with
\begin{equation}
	\dot{O}_{\bar{A}}
	= \bar{\cA} O_{\bar{A}} + O_{\bar{A}} \bar{\cA}^{\dagger} -i \cU a^{\st} \phi \cU^{\dagger} + i \cU \phi a^{\st} \cU^{\dagger}.
\end{equation}
Inserting factors of $\cU^{\dagger} \cU$ in the last two terms, this becomes
\begin{equation}
	\dot{O}_{\bar{A}}
	= \bar{\cA} O_{\bar{A}} + O_{\bar{A}} \bar{\cA}^{\dagger} 
	- i \left[ \cU a^{\st} \cU^{\dagger}, O_{\bar{A}} \right]
\end{equation}
and plugging in the definition of $\bar{\cA}$ gives
\begin{equation}
	\dot{O}_{\bar{A}}
	= i \left[ \bar{a} - \cU a^{\st} \cU^{\dagger}, O_{\bar{A}} \right]
	+ i \int_0^{\infty} e^{-\epsilon t} \left[ e^{i K_{\bar{A}} t} \dot{K}_{\bar{A}} e^{-i K_{\bar{A}} t}, O_{\bar{A}}\right].
\end{equation}
Since $\dot{K}_{\bar{A}}$ commutes with $K_{\bar{A}},$ and $\dot{K}_{A}$ commutes with $O_{\bar{A}}$, we may rewrite this in terms of the full modular Hamiltonian $h = K_{\bar{A}} - K_A$ as
\begin{equation}
	\dot{O}_{\bar{A}}
	= i \left[ \bar{a} - \cU a^{\st} \cU^{\dagger}, O_{\bar{A}} \right]
	+ i \int_0^{\infty} e^{-\epsilon t} \left[ e^{i h t} \dot{h} e^{-i  h t}, O_{\bar{A}}\right].
\end{equation}
This exactly matches our fundamental equation \eqref{eq:operator-flow-intro} after relabeling the modular zero mode and substituting $e^{i h t} = \Delta^{-it}.$

\section{Discussion}
\label{sec:discussion}
Bulk reconstruction for general entanglement wedges is an interesting open problem in AdS/CFT. Existing methods, such as the Petz map, involve Euclidean path integral constructions, and do not have a simple, manifestly Lorentzian representation. In this paper, we set up a framework for relational bulk reconstruction. Instead of reconstructing a particular code subspace operator, the goal here was to relate the operator reconstructions between two different code subspaces which are smoothly connected to each other by a one-parameter family of code subspaces which are all protected from erasure of the same boundary subregion. A particularly nice situation is when the maximally mixed state on the code subspace is everywhere full-rank with respect to $\bar{A}$ (i.e., cyclic and separating) along the above one-parameter family. This happens to be true in a class of ``measurement-based'' code subspaces in holography. In this case, we derived a flow equation for the reconstructed operator, which can in principle be integrated to relate the reconstructed operator at one point in the flow to another. This flow equation involved modular flow for the maximally mixed state on the code subspace, and gives a manifestly Lorentzian, relational formula for bulk reconstruction. We also noted that this formula resembles infinite-time limit of the Connes cocycle flow, and we gave some plausibility arguments for this connection contingent upon the late-time decay of modular flowed correlation functions. Finally, the same results were re-derived from a different point of view in the context of factorized Hilbert spaces by using modular reflections.    

We end with a discussion of some open questions and directions for future work:
\begin{itemize}

\item The connection between our formulas for relational bulk reconstruction and the infinite-time Connes cocycle is very suggestive, but it would be good to make this connection precise. In our attempt at making this connection rigorous, we encountered the condition stated in equation \eqref{eq:uniform-decay-condition}. This condition is naturally related to the decay of late-time modular flowed correlators. The connection between quantum chaos and ergodicity of modular flow, recently explored in \cite{Gesteau:chaos, Ouseph:chaos}, may be the underlying physics behind this. 

\item The late-time behavior of the cocycle appeared in \cite{Ceyhan:2018zfg} in the context of shape deformations of relative entropies.
It would be interesting to see if the techniques in that paper can lead to a more geometric understanding of our formulas for relational bulk reconstruction involving shape deformations of the bulk entanglement wedge \cite{Faulkner:2017tkh, Lewkowycz:2018sgn}.

 \item One disadvantage is that our formulas are quite abstract. For instance, it is very difficult to calculate modular flow for the maximally mixed state on most code subspaces of interest. We feel this is a shortcoming, and any explicit calculation would constitute significant progress. Determining the settings in which our formalism is valid by computing reflection operators and flow-reconstructed code subspace operators in toy models of holography might suggest a way forward.
 
\item It would be good to generalize our formalism away from the infinite-$N$ limit by relaxing the assumption of exact quantum error correction. This would also presumably allow one to make a connection between our formulas and the ones used for relational bulk reconstruction in \cite{Almheiri:2017fbd, Levine:2020upy, Engelhardt:2021mue}. The framework of asymptotically isometric codes introduced in \cite{Faulkner:2022ada} may be a good setting to consider for this purpose. 

\item The holographic code subspaces used in this work bear a striking resemblance to the ``probe black hole'' setting of \cite{Lamprou:probes1, Lamprou:probes2, Lamprou:probes3}. It would be good to understand this connection more precisely.

\end{itemize}

\subsection*{Acknowledgments}
We thank Abhijit Gadde, Gautam Mandal, Shiraz Minwalla and Sandip Trivedi for helpful discussions and Tom Faulkner for helpful comments on an earlier version of this draft. We are grateful to the long term workshop YITP-T-23-01 held at YITP, Kyoto University, where part of this work was done.
JS is supported by the AFOSR under award number FA9550-19-1-0360, by the DOE Early Career Award number DE-SC0021886, and by the Heising-Simons Foundation.

\appendix 

\section{Replica trick, modular reflections and bulk reconstruction} \label{sec:app}
In this appendix, we give a replica trick construction for the modular reflection operator. We then use this replica trick together with a Lewkowycz-Maldacena \cite{lewkowycz2013generalized} style argument to show that the modular reflection operator for the maximally entangled state on the code subspace implements bulk reconstruction in holography. This latter argument works the same way as the replica trick for the Petz map \cite{penington2022replica}. 
\subsection{Replica trick} \label{subsec:replica-trick}

We begin by discussing a replica trick approach for computing the modular reflection operator.  We will motivate this replica trick in the factorized setting. Let the state $\Psi$ be a Euclidean path integral state, where the path integral is performed on the manifold $I \times \Sigma$, with $I$ being a Euclidean time interval. We can think of this path integral as preparing a state in the Hilbert space $\cH \otimes \cH^\st$, where $\cH$ is the Hilbert space on one copy of $\Sigma$. In this case, recall from equation \eqref{eq:MROFacdef} that the reflection operator takes the form 
\beq 
\cU: \cH^{\st} \to \cH^{\st},\;\;\cU = \sum_{n} |\tch_{n}\rangle \langle \ch^{\st}_{n}|,
\eeq 
where we have neglected to write the degeneracy indices for simplicity. Consider matrix elements of this operator in some basis states for $\cH^{\st}$:
\beq
\langle j^\st |\cU |i^\st\rangle = \sum_n \langle j^\st | \tch_n\rangle \langle \ch_n^\st | i^\st\rangle.
\eeq 
We would like a path-integral description for these matrix elements. The path integral for the original state allows us to compute the following overlap:
\beq  \label{statePI}
\langle i, j^\st | \Psi\rangle = \sum_n \sqrt{p}_n \,\langle i |\ch_n\rangle \langle j^\st | \tch_n\rangle,
\eeq
where $|i,j^{\st}\rangle = |i\rangle \otimes |j^{\st}\rangle$ is a basis element in $\cH \otimes \cH^\st$. Since
\beq 
\langle \ch_n^\st | i^\st\rangle = \langle i| \ch_n\rangle,
\eeq 
we note that the matrix elements of the modular reflection operator are closely related to the path-integral for the original state, except for the factors of $\sqrt{p_n}$ in equation \eqref{statePI}. This motivates us to define a R\'enyi version of the modular reflection operator:
\beq 
\cU_{(n)} = \sum_n p_n^{n+\frac{1}{2}} |\tch_n\rangle \langle \ch_n^\st|,
\eeq
which has the matrix elements:
\beq
\langle j^\st |\cU_{(n)} |i^\st\rangle = \langle i,j^\st|\rho^{n}|\Psi\rangle.
\eeq 
In the $n\to -1/2$ limit, the matrix elements reduce to those of the modular reflection operator. Therefore, the path integral for $\cU_{(n)}$ is precisely the same as the path integral for $\rho^n \Psi$, but where we slice the path integral along a time coordinate which runs from the left boundary to the right boundary. This gives us a simple path-integral representation for the R\'enyi version of the modular reflection operator. To obtain the modular reflection operator, we compute the path-integral at integer $n$, and find an analytic continuation to $n = -1/2$. 

\begin{figure}
\centering
\includegraphics[width = 0.5\linewidth]{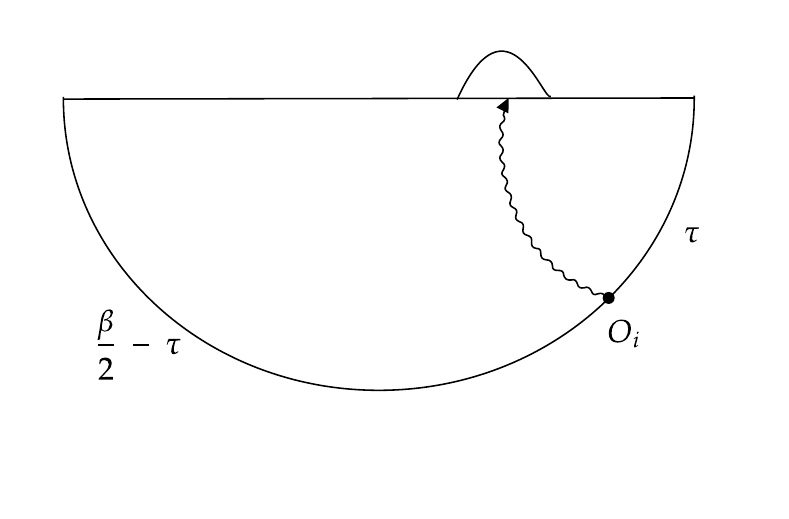}\caption{A state prepared by an operator insertion in Euclidean path integral.}
\label{fig:State}
\end{figure} 

\begin{figure}
\centering
\includegraphics[width=0.5\linewidth]{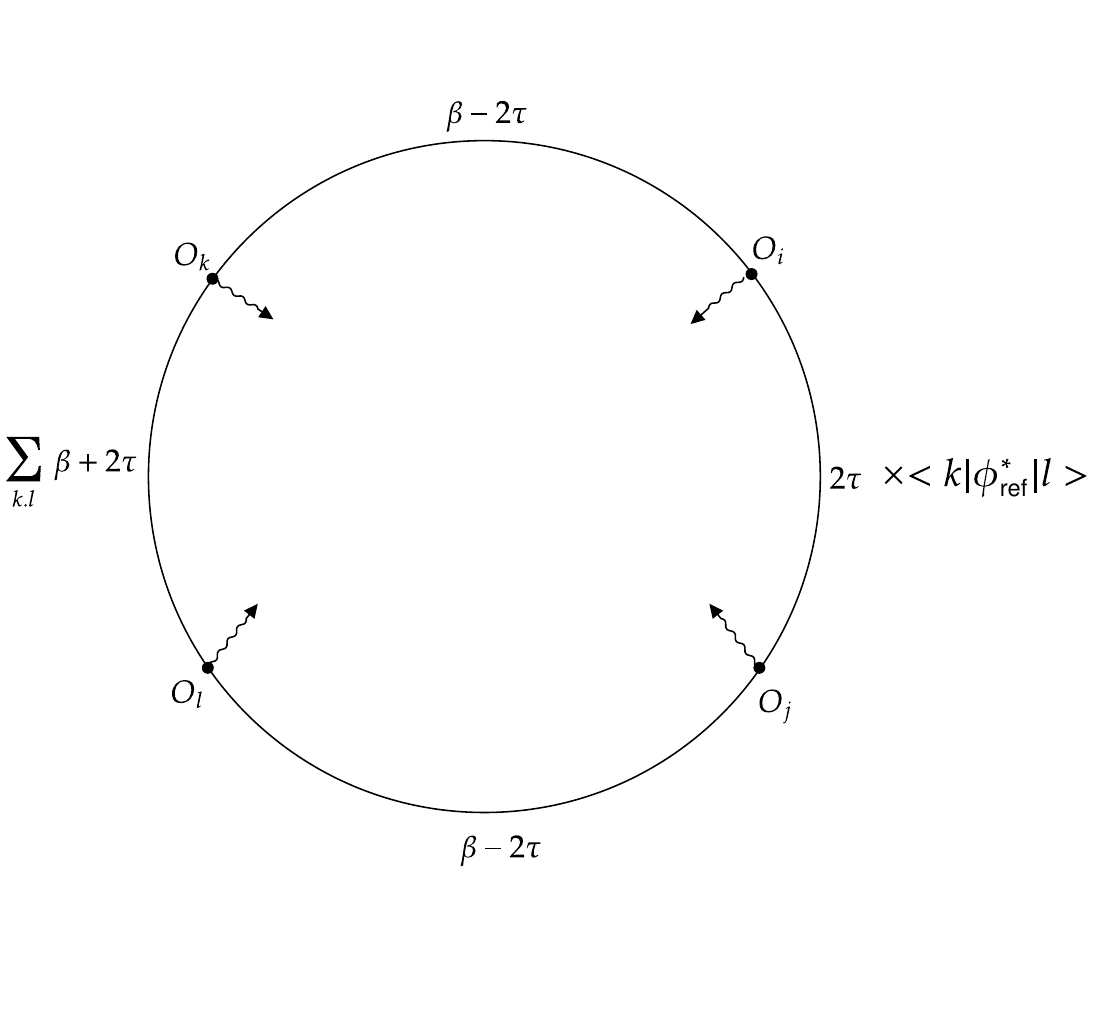}\caption{The replica path integral for computation of the matrix element of operator $O^{(n)}_\Abar$ in equation (\ref{eq:reflop}) for $n=1$. Ignoring the backreaction of the excitation, the bulk has $\mathbb{Z}_4$ replica symmetry. }
\label{fig:Replica}
\end{figure}

\subsection{Bulk reconstruction from replica trick}
The above replica trick can be used in the context of holography to see why the modular reflection operator implements bulk reconstruction. This argument is identical to that of \cite{penington2022replica} for the Petz recovery map, but we present it here for completeness. In order to use the replica trick, we first replace $\cU$ with its replica version $\cU_{(n)}$. Thus, consider the replicated operator
\beq\label{eq:reflop}
O_{\bar{A}}^{(n)} = \cU_{(n)}\, (\phi_{\text{ref}^\st}\otimes \mathbb{1}_{A^\st})\,\cU_{(n)}^{\dagger}.
\eeq 
The advantage of doing this, as before, is that we have an explicit path-integral construction for this replicated operator for integer $n$, and we can simply compute its matrix elements
\beq
\langle \psi_i | O_{\bar{A}}^{(n)} | \psi_j\rangle
\eeq 
between any two code-subspace states. We do this as follows: let the code subspace states $\psi_i$ be Euclidean path integral states. We imagine that the $i$th state is created by a particular operator insertion (labelled by the index $i$) or a source deformation in the Euclidean path integral (see figure \ref{fig:State}). The matrix element above can then be represented by the Euclidean path integral shown in figure \ref{fig:Replica}. In AdS/CFT, we think of this as providing the asymptotic boundary condition, and seek to fill it in with the minimum action solution of the bulk gravitational equations of motion. This is difficult in general, but we can make progress by invoking the Lewkowycz-Maldacena (LM) assumption which goes as follows: the boundary path integral, ignoring the code subspace insertions, has a $\mathbb{Z}_{2n+2}$ symmetry. In fact, this is the same boundary path integral (modulo the code subspace insertions) which computes the $(2n+2)$-th R\'enyi entropy in any one of the states $|\psi_i\rangle$, which we are assuming are all dual to the same bulk geometry (let's call it $M_1$), but only differ in the state of the bulk code subspace. The LM assumption is that the dominant bulk geometry is also $\mathbb{Z}_{2n+2}$-symmetric. This assumption makes the $n\to -\frac{1}{2}$ limit tractable. We note that there are two sources of the $n$-dependence: (i) the explicit dependence in $n$ coming from the fact that there are $(2n+2)$ ``slices of the pizza'' in the bulk, and (ii) from the $n$-dependence in the bulk geometry. In the $n\to -\frac{1}{2}$ limit, the geometry of each slice of the pizza becomes that of $M_1$. For instance, the conical deficit at the fixed point of the $\mathbb{Z}_{2n+2}$ symmetry is given by
\beq 
\theta = \frac{2\pi}{2n+2},
\eeq 
which in the $n\to -\frac{1}{2}$ limit approaches $2\pi$. The gravitational contribution is now simply $e^{-S_{\text{grav}}(M_1)}$, i.e., the gravitational on-shell action of $M_1$. Thus, in the $n\to -\frac{1}{2}$ limit, the calculation reduces to a bulk quantum field theory calculation on the fixed geometry $M_1$. In particular, this calculation corresponds to another quantum error correcting code, namely where we regard the qudit Hilbert space as a subspace of the bulk QFT Hilbert space. The analogous ``reconstructed'' operator on the entanglement wedge $\bar{a}$ of $\bar{A}$ is given by $o_{\bar{a}}= \cR_{\text{bulk}}\,(\phi_{\text{ref}^\st}\otimes \mathbb{1}_{a^\st})\,\cU_{\text{bulk}}^{\dagger}$, where now $\cU_{\text{bulk}}$ is the modular reflection operator of the maximally entangled state with respect to the bulk subregion $\text{ref}\cup a$. The Lewkowycz-Maldacena argument shows that the CFT calculation of the code subspace matrix elements of the operator $O_{\bar{A}}^{(n)}$ precisely lands us on the corresponding bulk QFT calculation involving $o^{(n)}_{\bar{a}}$. But since the bulk qubit was localized in the entanglement wedge $\bar{a}$ by construction, it must be decoupled from $a$. Therefore, we conclude that   
\beq
\lim_{n\to -\frac{1}{2}}\langle \psi_i | O_{\bar{A}}^{(n)} | \psi_j\rangle = \phi_{ij}. 
\eeq 

\section{Algebraic version of modular reflections}
\label{sec:appB}

In section \ref{sec:modular-reflections}, we defined the modular reflection operator for a quantum state on a factorized Hilbert space.
It is also possible to define the modular reflection operator for quantum states where subsystems are described not as tensor factors, but as general von Neumann algebras.
This is the setting of relevance to complementary subregions within a connected spatial slice in continuum QFTs, or to large $N$ limits in holographic conformal field theories. 

In the non-factorized setting, our starting point is a vector $\ket{\Psi}$ in a global Hilbert space $\cH,$ together with a von Neumann algebra $\alg$ which acts on the Hilbert space.\footnote{For background on von Neumann algebras, see \cite{Witten:2018zxz} or \cite{Sorce:2023fdx}.}
The vector $\ket{\Psi}$ defines an algebraic state on $\alg$, which is a map $\omega_{\Psi} : \alg \to \mathbb{C}$ given by
\beq
    \omega_{\Psi}(a) = \langle \Psi | a | \Psi \rangle, \qquad a \in \alg.
\eeq
We will assume that $\ket{\Psi}$ is \textit{separating} for $\alg,$ meaning that the only operator $a \in \alg$ satisfying $\omega_{\Psi}(a^{\dagger} a) = 0$ is the operator $a=0.$
We made an analogous assumption in the factorized setting by assuming that $\ket{\Psi}$ was full rank with respect to $A$.

The GNS Hilbert space corresponding to $\omega_{\Psi}$, denoted $\cH_{\text{GNS}},$ is defined to be the completion of the vector space $\alg$ with respect to the inner product\footnote{If $\ket{\Psi}$ were not separating, then this inner product would have null states, and one would have to take a quotient by null states to produce a Hilbert space.}
\beq
    \langle a | b \rangle_{\text{GNS}}
        = \omega_{\Psi}(a^{\dagger} b).
\eeq
The algebra $\alg$ has a natural representation $\rho$ on the GNS Hilbert space given by
\begin{equation}
    \rho(a) \ket{b}_{\text{GNS}} = \ket{ab}_{\text{GNS}}.
\end{equation}
We may think of $\rho$ as providing a natural isomorphism from $\alg,$ which acts on $\cH,$ to an algebra acting on $\cH_{\text{GNS}}.$
For more background on GNS Hilbert spaces and the properties of this natural isomorphism, see e.g. \cite[chapter 7]{conway2000course}.

In $\cH_{\text{GNS}},$ there is a special state corresponding to the identity operator, which we denote by $|\Psi^\st\rangle = |\mathbb{1} \rangle_{\text{GNS}}.$
The expectation values of operators in this state satisfy
\begin{equation}
    \langle \Psi^{\st} | \rho(a) | \Psi^{\st} \rangle
        = \omega_{\Psi}(a)
        = \langle \Psi | a | \Psi \rangle.
\end{equation}
In this sense, $\ket{\Psi^{\st}}$ and $\ket{\Psi}$ are two purifications of the single algebraic state $\omega_{\Psi}.$
It is then natural to define the modular reflection operator $\cR: \cH_{\text{GNS}} \to \cH$ by the formula
\begin{equation}
    \cR \rho(a) \ket{\Psi^{\st}} = a \ket{\Psi}.
\end{equation}
This map is easily seen to be isometric via the following computation:
\begin{equation}
    \langle \cR \rho(a) \Psi^{\st}|\cR \rho(b) \Psi^{\st} \rangle
        = \langle a \Psi | b \Psi \rangle
        = \langle \rho(a) \Psi^{\st} | \rho(b) \Psi^{\st} \rangle.
\end{equation}
Note also that vectors of the form $\rho(a) \ket{\Psi^{\st}} = \ket{a}_{\text{GNS}}$ are dense in $\cH_{\text{GNS}},$ since $\cH_{\text{GNS}}$ was defined to be the Hilbert space completion of $\alg.$
Consequently, the operator $\cR$ can be extended by continuity to all of $\cH_{\text{GNS}}.$
This extension is what we call the \textit{modular reflection operator} in the algebraic setting.

In the factorized setting, the modular reflection operator had two important properties: it was an isometry, and it acted only on the ``second factor'' $\cH_{A}^{\st}$ in the canonical purification.
In the current setting, $\cR$ is still an isometry, but the GNS space $\cH_{\text{GNS}}$ does not factorize.
However, there is still a sense in which $\cR$ acts only on the complement of $\alg,$ which is that it commutes with all operators in $\alg$:
\beq 
b \,\cU\, \rho(a) |\Psi^\st\rangle = b\, a|\Psi\rangle = \cU \,\rho(b)\, \rho(a) |\Psi^\st\rangle.
\eeq 

In section \ref{sec:modular-reflections}, we derived a flow equation for the modular reflection operator along a one-parameter family of full-rank states, which lets us relate the modular reflection operator for some initial state $\Psi_0$ to the final state $\Psi_1$. In the algebraic setting, this gets a bit confusing because the domain of the reflection operator is the Hilbert space $\cH_{\text{GNS}}$ defined with respect to the state $\omega_{\Psi_{\lambda}}$, and for different states along the flow we will get different Hilbert spaces. There is one simplified context in which this can be made sense of, which is when the reduced state $\omega_{\Psi_{\lambda}}$ on $\alg$ is $\lambda$-independent. In this case, the domains are all given by the same Hilbert space, and we can meaningfully compare the reflection operators at two different values of $\lambda$. In such a situation, we can define
\beq 
U_{\Psi_0|\Psi_1}: \cH \to \cH,\;\; U_{\Psi_0|\Psi_1} = \cR_{\Psi_1}\cdot \cR^{\dagger}_{\Psi_0}.
\eeq
In the holographic context, if $\Psi_0$ and $\Psi_1$ are the maximally entangled states corresponding to two code subspaces protected against the erasure of the region $A$, then the map $U_{\Psi_0|\Psi_1}$ relates the corresponding reconstructed operators:
\beq 
O_{\bar{A}}(1) = U_{\Psi_0|\Psi_1}\, O_{\bar{A}}(0)\,U_{\Psi_0|\Psi_1}^{\dagger}. 
\eeq

\bibliographystyle{JHEP}
\bibliography{bibliography}

\end{document}